\documentclass[twocolumn]{aastex631}

\usepackage{physics}
\usepackage{graphicx}
\usepackage{booktabs}

\begin{document}

\title{Very Massive Stars and High N/O: A Tale of the Nitrogen-enriched Super Star Cluster \\ in the Sunburst Arc}

\correspondingauthor{Yanlong Shi}
\email{yanlong@cita.utoronto.ca}

\newcommand{\cita}{Canadian Institute for Theoretical Astrophysics, University of Toronto, Toronto, ON M5S 3H8, Canada}
\newcommand{\ucbphys}{Department of Physics, University of California, 366 Physics North MC 7300, Berkeley, CA 94720, USA}
\newcommand{\ucbast}{Department of Astronomy, University of California, 501 Campbell Hall \#3411, Berkeley, CA 94720, USA}
\newcommand{\uoftastro}{David A. Dunlap Department of Astronomy \& Astrophysics, University of Toronto, 50 St. George St., Toronto, ON M5S 3H4, Canada}

\author[0000-0002-0087-3237]{Yanlong Shi}
\affiliation{\cita}

\author[0000-0003-2091-8946]{Liang Dai}
\affiliation{\ucbphys}

\author[0000-0002-8659-3729]{Norman Murray}
\affiliation{\cita}

\author[0000-0001-9582-881X]{Claire S. Ye}
\affiliation{\cita}

\author[0000-0001-9732-2281]{Christopher D. Matzner}
\affiliation{\uoftastro}

\author[0000-0002-2282-8795]{Massimo Pascale}
\affiliation{\ucbast}

\begin{abstract}

The lensed Sunburst Arc ($z = 2.369$) hosts a young ($\sim2$--$4\,\rm Myr$), massive ($M_\star \sim 10^7\,M_\odot$), compact ($R_{\rm eff} \sim 8\,\rm pc$) Lyman-continuum (LyC) leaking super star cluster, which powers a compact ($< 10\,\rm pc$), high-pressure nebula at sub-solar metallicity $\sim0.2\,Z_\odot$ and with an anomalously elevated nitrogen-to-oxygen ratio $\log({\rm N/O}) \sim -0.2$. We present semi-analytic models and 3D magnetohydrodynamic simulations in an attempt to reproduce this system. The results indicate that the progenitor giant molecular cloud (GMC) may have $M_{\rm cloud} \gtrsim 3 \times 10^7\,M_\odot$ and $R_{\rm cloud} \sim 70\,\rm pc$, corresponding to a surface density $\sim10^3$--$10^4\,M_\odot\,{\rm pc}^{-2}$. Incorporating feedback from individual very massive stars (VMSs; $\ge 100\,M_\odot$) sampled from the Kroupa initial mass function, we find that their winds rapidly enrich $\sim 10^4\,M_\odot$ of nearby gas with nitrogen ($\sim 1\,$dex) and helium ($\sim 0.1$--$0.2\,$dex). In the first $1$--$3\,$Myr, some cold gas falls to the system center where a central cluster builds up from sub-cluster mergers. There, the gas is photoionized, pressurized, and chemically enriched by the newly formed VMSs, before being radiatively expelled in the next $\sim1\,\rm Myr$. We find that both VMS feedback and a high-surface-density progenitor GMC are necessary to reproduce the observed nebular properties, such as high N/O, high pressure, and stellar proximity. Low metallicity ($Z \le 0.004$) may be essential to avoid overproduction of carbon from WC stars. Such enrichment processes localized to compact starbursts may have caused strong nitrogen emission from dense ionized gas as observed in high-redshift galaxies such as GN-z11 and GS\_3073.

\end{abstract}

\keywords{Young star clusters (1833), Chemical enrichment (225), Star formation (1569), Massive stars (732)}

\section{Introduction} \label{sec:intro}

Strong gravitational lensing enables a magnified view of distant galaxies. The Sunburst Arc at $z\approx 2.369$ is one of the brightest lensed galaxies~\citep{DahleAghanimGuennou_2016A&A...590L...4D}. Inside that galaxy is a compact star-forming clump that features a high Lyman continuum (LyC) escape fraction~\citep{Rivera-ThorsenDahleGronke_2017A&A...608L...4R,Rivera-ThorsenDahleChisholm_2019Sci...366..738R}, which is determined to be a $\sim 2$--$4\,{\rm Myr}$-old~\citep{ChisholmRigbyBayliss_2019ApJ...882..182C,PascaleDaiMcKee_2023ApJ...957...77P,MestricVanzellaUpadhyaya_2023A&A...673A..50M,Rivera-ThorsenChisholmWelch_2024A&A...690A.269R} super star cluster with a stellar mass in excess of $\sim 10^7\,M_\odot$~\citep{PascaleDaiMcKee_2023ApJ...957...77P,Rivera-ThorsenChisholmWelch_2024A&A...690A.269R} and an effective radius of $\sim 8\,\rm pc$~\citep{VanzellaCastellanoBergamini_2022A&A...659A...2V,MestricVanzellaUpadhyaya_2023A&A...673A..50M}. The gas metallicity is $Z\sim 0.2\,Z_\odot$~\citep{PascaleDaiMcKee_2023ApJ...957...77P,Rivera-ThorsenChisholmWelch_2024A&A...690A.269R,WelchRivera-ThorsenRigby_2025ApJ...980...33W}. Moreover, \citet{PascaleDaiMcKee_2023ApJ...957...77P} found through nebular spectroscopy highly-pressurized ($p/k_{\rm B} \gtrsim 10^9\,\rm K\, cm^{-3}$) photo-ionized gas within $\sim 10\,$pc of the cluster center, which features a high nitrogen abundance $\log (\rm N/O) \sim -0.21$, while the outer, lower-pressure gas is not enriched. This is in strong excess of the median ISM N/O value for its metallicity according to empirical relations~\citep{NichollsSutherlandDopita_2017MNRAS.466.4403N}.

Similar nitrogen abundance anomalies have been observed from systems at various cosmic redshifts~\citep{Marques-ChavesSchaererKuruvanthodi_2024A&A...681A..30M}. Remarkably, several high-redshift galaxies~\citep[especially GN-z11 and GS\_3073, e.g.,][and more references in \S~\ref{sec:discussion}]{BunkerSaxenaCameron_2023A&A...677A..88B,CameronKatzRey_2023MNRAS.523.3516C,JiUblerMaiolino_2024MNRAS.535..881J} and AGNs~\citep{IsobeMaiolinoD'Eugenio_2025MNRAS.541L..71I} show nitrogen-enriched nebulae in JWST observations. Compared with the Sunburst Arc super star cluster, the nitrogen enhancement for some of these galaxies is more extreme~\citep[$\log(\rm N/O) \gtrsim 0.5$, e.g.,][]{JiUblerMaiolino_2024MNRAS.535..881J}.  Even in the Sunburst Arc, another compact young star cluster is found to host an N-enriched, ultra-dense ($\log n_{\rm e}=7$--$8$) nebula~\citep{PascaleDai_2024ApJ...976..166P}~\citep[also see][]{ChoeEmilRivera-ThorsenDahle_2025A&A...698A..16C}. Toward low redshifts, nitrogen emitters appear exceedingly rare~\citep{BhattacharyaKobayashi_2025arXiv250811998B}, but the central young starburst of the blue compact dwarf galaxy Mrk 996 is one local example with a very dense enriched nebula~\citep{JamesTsamisBarlow_2009MNRAS.398....2J}. The dwarf galaxy NGC 5253 hosts another $\sim 10^5\,M_\odot$ embedded super star cluster with localized nitrogen enrichment~\citep{SmithCrowtherCalzetti_2016ApJ...823...38S}~\citep[also see][]{Abril-MelgarejoJamesAloisi_2024ApJ...973..173A,
PruijtMonreal-IberoWeilbacher_2025arXiv250907810P}.

Several stellar sources of pollution have been proposed to explain the nitrogen enrichment, including Wolf-Rayet (WR) stars~\citep{KobayashiFerrara_2024ApJ...962L...6K,FukushimaYajima_2024PASJ...76.1122F,FluryArellano-CordovaMoran_2025MNRAS.543.3367F}, temporally differential winds that are enriched by core-collapse supernovae~\citep[][]{RizzutiMatteucciMolaro_2025A&A...697A..96R} or asymptotic giant branch (AGB) stars~\citep{D'AntonaVesperiniCalura_2023A&A...680L..19D,McClymontTacchellaSmith_2025arXiv250708787M,BhattacharyaKobayashi_2025arXiv250811998B}, OB stars~\citep{TapiaBekkiGroves_2024MNRAS.534.2086T}, Very Massive Stars~\citep[VMSs;][]{Vink_2023A&A...679L...9V}, and Supermassive Stars~\citep{CharbonnelSchaererPrantzos_2023A&A...673L...7C,NandalWhalenLatif_2025ApJ...994L..11N,GielesPadoanCharbonnel_2025MNRAS.544..483G}. An intriguing observed fact is that these nitrogen enriched nebulae always exhibit unusually high electron density $\log n_{\rm e}\gtrsim 5$~\citep{JamesTsamisBarlow_2009MNRAS.398....2J,PascaleDaiMcKee_2023ApJ...957...77P, JiUblerMaiolino_2024MNRAS.535..881J, ToppingStarkSenchyna_2024MNRAS.529.3301T,YanagisawaOuchiWatanabe_2024ApJ...974..266Y,IsobeMaiolinoD'Eugenio_2025MNRAS.541L..71I}, which hints at gaseous environments subject to strong radiative or hydrodynamic compression in the vicinity of young massive stars~\citep{PascaleDaiMcKee_2023ApJ...957...77P}.  

Given the potential connections between high-redshift nitrogen-enriched galaxies and globular cluster stars with similar abundance anomalies~\citep{IsobeOuchiTominaga_2023ApJ...959..100I,JiBelokurovMaiolino_2025MNRAS.tmp.1982J}, understanding the LyC cluster of the Sunburst Arc at Cosmic Noon can shed crucial light on the very high-$z$ nitrogen-enriched galaxies. In particular, the LyC cluster is $\gtrsim 100$ times more massive than known young massive clusters of the Milky Way and the Magellenic Clouds. For its young age, a rapid mechanism of enrichment is required~\citep{PascaleDaiMcKee_2023ApJ...957...77P,Rivera-ThorsenChisholmWelch_2024A&A...690A.269R}.

In this study, we focus on the scenario associated with VMSs~\citep{Vink_2023A&A...679L...9V}. VMSs are massive ($\gtrsim 100\, M_\odot$) and short-lived ($2-3\,\rm Myr$) stars~\citep{VinkHegerKrumholz_2015HiA....16...51V} that are beyond the normally assumed mass range of an initial mass function (IMF)~\citep[e.g.,][]{Kroupa_2002Sci...295...82K}. Their abundance may be enhanced with a top-heavy IMF, which is theoretically expected in the dense star-forming environments of globular cluster progenitors~\citep{Matzner_2024ApJ...975L..17M}. Unlike less massive stars, VMSs can shed order unity of their initial mass through winds on the main sequence~\citep{VinkMuijresAnthonisse_2011A&A...531A.132V,Vink_2018A&A...615A.119V,SabhahitVinkHiggins_2022MNRAS.514.3736S}. Moreover, they are typically rotating and highly convective, making the surface abundance of elements different from that of less massive stars~\citep{HigginsVinkHirschi_2023MNRAS.526..534H,HigginsVinkHirschi_2024MNRAS.533.1095H}. In fact, they are often spectroscopically classified as WNh stars for prominent emission features from their strong winds. As such, VMS winds can be a plausible source of element enrichment for the early phase of chemical evolution of the ISM~\citep{LahenNaabSzecsi_2024MNRAS.530..645L}. Given the young age of the Sunburst Arc LyC cluster, VMSs offer a rapid route to reach the high nitrogen abundance~\citep{Vink_2023A&A...679L...9V}.

Individual VMSs have already been found in the Local Group~\citep[e.g., in the R136 star cluster of the Large Magellanic Cloud,][]{CrowtherSchnurrHirschi_2010MNRAS.408..731C,KeszthelyiBrandsdeKoter_2025A&A...700A.186K}. However, nitrogen-rich dense ionized gas has not been found in this $\sim 10^5\,M_\odot$ cluster.
Through integral spectral features, VMSs are revealed in both low-redshift galaxies~\citep{SenchynaStarkCharlot_2021MNRAS.503.6112S,MartinsSchaererMarques-Chaves_2023A&A...678A.159M, WoffordSixtosCharlot_2023MNRAS.523.3949W} and in high-$z$ galaxies~\citep{UpadhyayaMarques-ChavesSchaerer_2024A&A...686A.185U}. Based on spectroscopic evidence of VMSs, \cite{SmithCrowtherCalzetti_2016ApJ...823...38S} suggest that they cause the observed nitrogen pollution in the vicinity of the $\sim 10^5\,M_\odot$ embedded nascent super star cluster in NGC 5253. There, nitrogen enhancement shows in the optical [N\textsc{ii}]$\lambda\lambda$6548,6584, but does not show as the FUV N\textsc{iii}]$\lambda$1750 multiplet that would reveal dense $\log n_e>4$ ionized gas.

Observational studies support the presence of VMSs in the Sunburst Arc LyC cluster. For example, \citet{MestricVanzellaUpadhyaya_2023A&A...673A..50M} found several key spectrum signatures of VMSs~\citep[especially He\textsc{ii}$\lambda1640$, also see][]{GrafenerVink_2015A&A...578L...2G,MartinsPalacios_2022A&A...659A.163M,MartinsPalaciosSchaerer_2025A&A...698A.262M} in the cluster. Moreover, the effective radius measured for LyC radiation is significantly smaller than that at $1700\,\rm \r{A}$, suggesting central segregation of these stars~\citep{MestricVanzellaUpadhyaya_2023A&A...673A..50M}. On the other hand, \citet{Rivera-ThorsenChisholmWelch_2024A&A...690A.269R} reported detection of the WR ``blue bump'' and weaker ``red bump,'' and generally attribute them to WNh stars, but a VMS population is also desired. 

Despite a number of previous works that explain the nitrogen enrichment of the Sunburst Arc LyC cluster, direct simulations on cluster formation, gas dynamical evolution, and chemical self-enrichment are required to reveal details of this process. In this study, we follow the method as presented in \citet{Shi_inprep} to run simulations that integrate FIRE-3 star-formation physics~\citep{HopkinsKeresOnorbe_2014MNRAS.445..581H,HopkinsWetzelKeres_2018MNRAS.480..800H,HopkinsWetzelWheeler_2023MNRAS.519.3154H} and the evolution of individual VMSs, allowing us to explicitly model VMS feedback. Here, we further implement stellar evolution tracks from the PARSEC dataset~\citep{NguyenCostaGirardi_2022A&A...665A.126N,CostaShepherdBressan_2025A&A...694A.193C} such that chemical yields from VMSs are included in the simulation.

This article is organized as follows. In \S~\ref{sec:toy_model}, we compile PARSEC data as a proof of concept to show that VMSs can account for the rapid nitrogen enrichment in young clusters such as the Sunburst Arc LyC cluster. We then describe our simulations: the methods in \S~\ref{sec:sim} and the results in \S~\ref{sec:res}, where we present fiducial runs reproducing the LyC cluster features and control runs exploring star formation and VMS feedback. Section~\ref{sec:discussion} discusses the implications for high-redshift nitrogen-enriched galaxies, and Section~\ref{sec:conclusion} summarizes our conclusions.

\section{Chemical enrichment by stellar winds: a toy model} \label{sec:toy_model}

Most chemical elements heavier than helium~\citep{AlpherBetheGamow_1948PhRv...73..803A} are forged in stars~\citep{BurbidgeBurbidgeFowler_1957RvMP...29..547B}. In this article, we implement sub-grid stellar yield models based on the public evolution tracks, PARSEC v2.0~\citep{NguyenCostaGirardi_2022A&A...665A.126N,CostaShepherdBressan_2025A&A...694A.193C}, which provides the full evolution history of non-rotating, intermediate to VMSs for a range of initial metallicity $10^{-11} - 0.03$~\citep[][]{CostaShepherdBressan_2025A&A...694A.193C}, largely covering the necessary parameter space required by our simulation.

Here, we present a simple semi-analytic toy model to describe the chemical enrichment of the ISM by stellar winds. We assume that the stellar population follows the Kroupa initial mass function $f(m)$~\citep{Kroupa_2002Sci...295...82K}. In particular, $f(m) \propto m^{-2.3}$ at the high-mass end, and we assume an upper mass cutoff defined as $m_{\rm max}$. We normalize $f(m)$ such that $\int_{m_{\rm min}}^{m_{\rm max} } f(m) \dd m = 1$ where $m_{\rm min}=0.01\, M_\odot$.

We weight each stellar evolution track to compute the cumulative ejecta mass of a stellar population over time. For a track with ZAMS mass $m_j \in [2\,M_\odot, m_{\rm max}]$, the corresponding mass bin is $[m_j^-, m_j^+] = [(m_{j-1}+m_{j})/2, (m_{j}+m_{j+1})/2]$. The weight is then $w_j = \int_{m_j^-}^{m_j^+} f(m) \dd m$. Assuming the initial mass of the stellar population is $M_\star$, the cumulative mass ejecta for element $i$ per unit mass, denoted as $\eta_i(t)$, is then
\begin{align}
    \eta_i(t) = \frac{1}{\langle m \rangle} \sum_j w_j \left[\int_0^t \dot m (t)\,X_i(t)\,\dd t\right]_j.
\end{align}
Here $\langle m \rangle$ is the mean stellar mass obtained from the IMF, textcolor{black}{$\dot m$ is the mass loss rate of the star}, and $X_i(t)$ is the surface abundance of element $i$. All these quantities are available from the PARSEC data.

We note that star formation is time-dependent, with the star formation rate (SFR) $\dot M_\star (t)$. Then the mass ejecta evolution should be convolved over this SFR:
\begin{align}
    M^i_{\rm ejecta} (t) = \int_0^t \dot M_\star (\tau)\,\eta_i (t-\tau) \,\dd \tau. \label{equ:ejecta_mass}
\end{align}

Simulations of star formation from giant molecular clouds (GMCs) suggest that the SFR peaks roughly at one free-fall time~\citep{GrudicHopkinsFaucher-Giguere_2018MNRAS.475.3511G,Shi_inprep}. As a toy model, we assume that the free-fall time is $t_{\rm ff}=1\,\rm Myr$ and the SFR has a Gaussian with a standard deviation of $0.75\,\rm Myr$. We also assume a high star formation efficiency of $0.4$. The star formation history is initialized with $M_\star = 0$ at $t = 0$. 

As star formation proceeds, the pristine gas is partly converted into stars and partly mixed with stellar ejecta. Simulations indicate that stellar feedback removes a large fraction of gas from clusters~\citep{GrudicHopkinsFaucher-Giguere_2018MNRAS.475.3511G}. Even in compact globular cluster progenitors such as the Sunburst Arc LyC cluster, radiative feedback can expel gas while producing LyC emission~\citep{MenonLancasterBurkhart_2024ApJ...967L..28M,MenonBurkhartSomerville_2025ApJ...987...12M}. In our toy model, gas depletion follows the star formation rate with a $1\,\rm Myr$ delay and is assumed highly efficient, leaving only 2\% of the initial GMC mass as gas at late times. Consequently, the mixed gas is dominated by stellar-wind ejecta, consistent with analytic estimates. For example, \citet{TapiaBekkiGroves_2024MNRAS.534.2086T} obtained $M_{\rm gas}/M_\star \sim 0.01$ at the time of the N/O anomaly, while our assumption of $M_{\rm gas}/M_\star = 0.02/0.4 = 0.05$ is slightly more conservative.

\begin{figure}
    \centering
    \includegraphics[width=\linewidth]{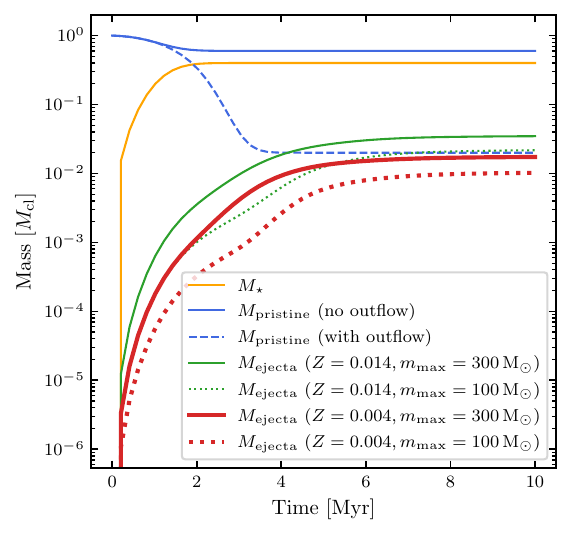}
    \caption{Mass evolution of star formation and feedback in the toy model (\S~\ref{sec:toy_model}), including the total mass of stars $M_\star$ and the mass of the pristine gas $M_{\rm pristine}$ (without or with feedback-driven outflow). Moreover, we show the cumulated mass of wind ejecta $M_{\rm ejecta}$ from stellar populations of different initial metallicity $Z$ and IMF cutoff $m_{\rm max}$. 
    }
    \label{fig:toy_model_mass}
\end{figure}

Figure~\ref{fig:toy_model_mass} shows the evolution of the total stellar mass and pristine gas mass, with and without outflow. Overlaid are the stellar-wind ejecta masses $M_{\rm ejecta}$, computed from Eq.~\eqref{equ:ejecta_mass} for different initial metallicities $Z$ and IMF cutoffs $m_{\rm cut}$ (with or without VMSs). For both $Z=0.014$ and $Z=0.004$, VMSs in the range $100$--$300\,M_\odot$ double the total wind ejecta. Assuming a star formation efficiency of $0.4$, about $10\%$ ($3\%$) of the stellar mass is lost in winds for $Z=0.014$ ($Z=0.004$), comparable to the remaining pristine gas under the feedback-driven outflow assumption.

\subsection{Abundances of pristine gas}

The elemental abundances of the pristine gas are dependent on the metallicity~\citep[e.g.,][and other citations below]{BergSkillmanHenry_2016ApJ...827..126B} due to prior chemical evolution in the host galaxy. They should be modeled in this toy model as well as in simulations. Knowing that crudely $Z\propto \rm O/H$, we use oxygen abundance as a proxy for metallicity. We define $x = 12+\log({\rm O/H})$, and for the solar metallicity defined in the FIRE simulations~\citep[following][]{WiersmaSchayeSmith_2009MNRAS.393...99W}, $x_\odot \approx 8.75$.

\begin{figure}
    \centering
    \includegraphics[width=\linewidth]{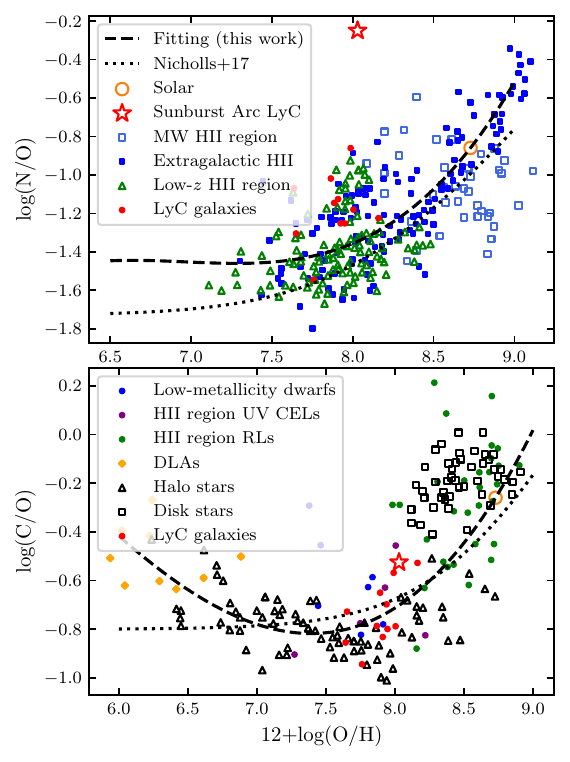}
    \caption{textcolor{black}{Fitting relations of log(N/O) (Eq.~\ref{equ:no_fitting}; \emph{top}) and log(C/O) (Eq.~\ref{equ:co_fitting}; \emph{bottom}), which match the solar metallicity used in FIRE \citep{WiersmaSchayeSmith_2009MNRAS.393...99W,HopkinsWetzelWheeler_2023MNRAS.519.3154H}. We also present data points for log(N/O) \citep[][and references therein]{MollaVilchezDiaz_2007astro.ph..1691M} and log(C/O) \citep[][and references therein]{BergSkillmanHenry_2016ApJ...827..126B}, where the low-redshift LyC leaking galaxies (red dots) are compiled from \citet{IzotovSchaererWorseck_2023MNRAS.522.1228I}.}
    }
    \label{fig:no_co_fittings}
\end{figure}

We fit the data presented in \citet{MollaVilchezDiaz_2007astro.ph..1691M} (galactic/extragalactic H{\sc ii} regions, low metallicity galaxies; see references therein) and force the relation to match FIRE's solar-metallicity ISM at $x=x_\odot$ (for compatibility with FIRE conventions), we have
\begin{align}
    \log({\rm N/O}) = -41.2+17.2\,x-2.48\,x^2+0.119\,x^3. \label{equ:no_fitting}
\end{align}
This relation asymptotically leads to $\log(\rm N/O)\approx-1.4$ at $Z\sim 0.2\, Z_\odot$. This is roughly in agreement with local observations~\citep{VincenzoBelfioreMaiolino_2016MNRAS.458.3466V}. Similarly, fitting data in \citet{BergSkillmanHenry_2016ApJ...827..126B}, we obtain 
\begin{align}
    \log({\rm C/O}) = -2.36+2.55\,x-0.606\,x^2+0.0391\,x^3. \label{equ:co_fitting}
\end{align}
This relation is not monotonic in $x$, which agrees with observations of halo stars~\citep{FabbianNissenAsplund_2009A&A...500.1143F}. 

textcolor{black}{Fig.~\ref{fig:no_co_fittings} presents the fitting relations (Eqs.~\ref{equ:no_fitting} and \ref{equ:co_fitting}) along with data from \citet{MollaVilchezDiaz_2007astro.ph..1691M} and \citet{BergSkillmanHenry_2016ApJ...827..126B}. Given the large scattering of the raw data, our fitting relations approximately match empirical relations in \citet{NichollsSutherlandDopita_2017MNRAS.466.4403N}. We also plot the log(N/O) and log(C/O) of the Sunburst Arc LyC cluster \citep{PascaleDaiMcKee_2023ApJ...957...77P}, which suggests anomalously elevated N/O, while C/O is not significantly elevated.}

\begin{figure*}
    \centering
    \includegraphics[width=\linewidth]{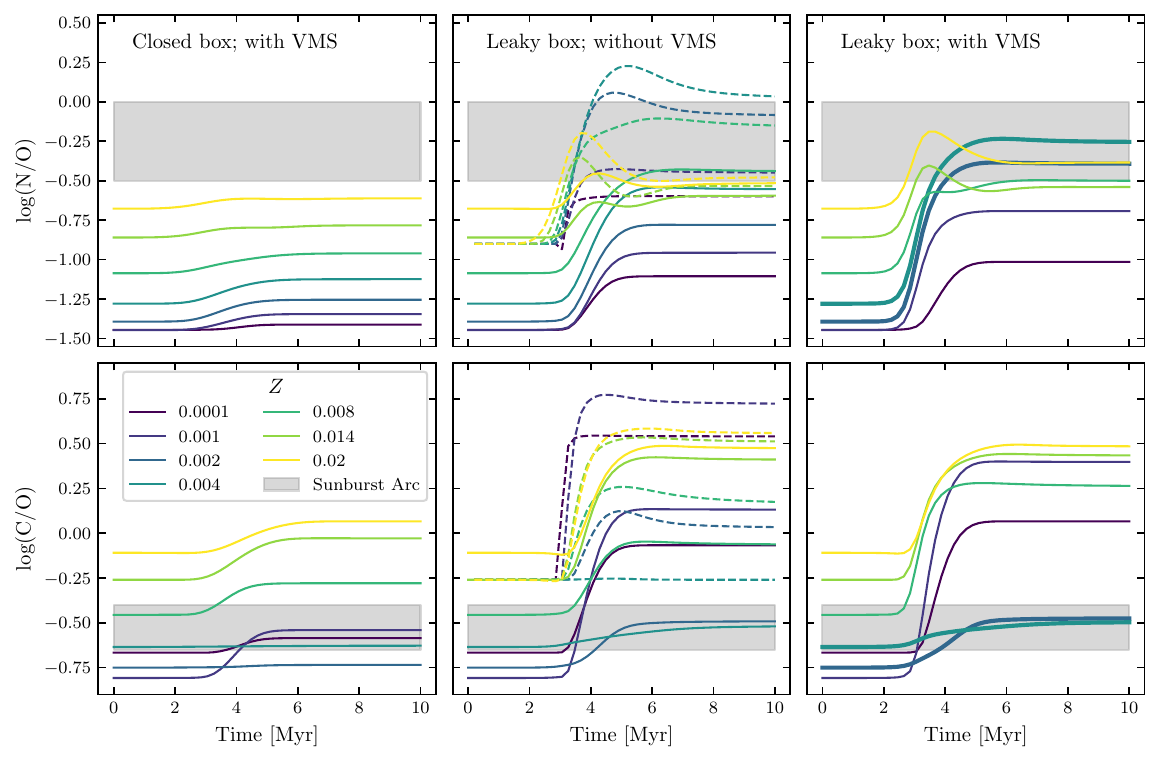}
    \caption{Evolution of the relative nitrogen (\emph{top}) and carbon (\emph{bottom}) of the interstellar medium in the toy model (\S~\ref{sec:toy_model}). In each panel, we show 6 tracks with various initial metallicities $Z$, and the rough estimated range of the Sunburst Arc cluster for reference~\citep[shaded regions,][]{PascaleDaiMcKee_2023ApJ...957...77P}. From left to right, in each column is the model with different assumptions (closed box or leaky box, with or without VMSs of $> 100\,M_\odot$). In particular, we also show the abundance ratios in the stellar wind (i.e., without mixing with the interstellar medium; \emph{dashed lines}).
    }
    \label{fig:toy_model_chemical}
\end{figure*}

\subsection{Closed box vs. leaky box}

Our toy model considers two scenarios for the pristine gas: one without outflow and one with feedback-driven outflow that removes a fraction of the gas from the cluster. These correspond to the classical ``closed-box'' and ``leaky-box'' models, respectively.

The chemical components of the polluted gas evolve due to enrichment from the wind ejecta. Fig.~\ref{fig:toy_model_chemical} shows the evolution of $\log(\rm N/O)$ and $\log (\rm C/O)$ at metallicities $Z=0.001, 0.002, 0.004, 0.008, 0.014, 0.02$, in comparison with the approximate observed ranges (\emph{gray shaded regions}) of the high-pressure gas of the LyC cluster taken from \citet{PascaleDaiMcKee_2023ApJ...957...77P}. For the closed box model, even if VMSs are included in the stellar population, none of the evolution tracks can reach the observed N/O.  For the leaky box model without the VMS population (middle column), only evolution tracks with $Z=0.014, 0.02$ can marginally reach the lower bound of the observed range at late times, but their $\log(\rm C/O)$ is well above the observed range. Even if we consider the extreme case that the leftover gas in the cluster is purely ejecta-dominated (\emph{dashed lines}), the C/O ratio is well above the observation.

The only case that can reproduce both nitrogen and carbon abundances is the leaky box model with the VMS population (right column). We find that evolution tracks with $Z=0.002,0.004$ are simultaneously in the observed range of $\log(\rm N/O)$ and $\log (\rm C/O)$ after $\sim 4\,\rm Myr$. Some other evolution tracks can reach the N/O anomaly but typically produce too much carbon after $\sim 3\,\rm Myr$.

The carbon abundance is another constraint to be satisfied. Based on it,  \citet{TapiaBekkiGroves_2024MNRAS.534.2086T} argues that WR stars should be excluded from the model to avoid over-production of carbon. However, the PARSEC model here suggests that carbon enrichment can be modest to match the observation for certain metallicities (i.e., $Z=0.002$ and $Z=0.004$ here). Interestingly, these are roughly $0.2\,Z_\odot$, the metallicity of the LyC cluster~\citep{PascaleDaiMcKee_2023ApJ...957...77P}. This further supports the ``leaky box + VMS model'' as a possible explanation for this specific system. 

textcolor{black}{It is worth mentioning the two low-metallicity tracks ($Z=0.001,10^{-4}$) also overproduce carbon after $\sim 3\,\rm Myr$. This is primarily due to the ``dredge-up events'' happening to a fraction of massive stars near the end of their time, where the convective envelope penetrates to the helium-burning core and brings the carbon to the surface \citep[Appendix~C of][]{CostaShepherdBressan_2025A&A...694A.193C}. This could affect the chemical feedback in systems with even lower metallicity than discussed here.
}

\subsection{Summary and caveats}

This toy model (Fig.~\ref{fig:toy_model_chemical}) suggests that winds from the stellar population can explain the N/O anomaly in the Sunburst Arc LyC cluster if these conditions are satisfied: 1) the inclusion of VMSs $\gtrsim 100\,M_\odot$ such that up to $\sim 10\%$ of the stellar mass can be ejected as stellar winds; 2) star formation efficiency should be high enough ($\gtrsim 0.1$); 3) initial metallicity of the GMC should be in the range of $Z\sim 0.002 - 0.004$; 4) a large fraction of the pristine gas should be removed from system by stellar feedback, so the remaining pristine gas is comparable with the wind ejecta in mass (Fig.~\ref{fig:toy_model_mass}); 5) the free-fall time of the GMC should be shorter than the age of the LyC cluster ($\lesssim4\,\rm Myr$), such that radiative feedback have already expelled most of gas at the time of nitrogen enrichment.

This toy model omits several key physical processes. First, it lacks a realistic star formation history, which affects the timing and strength of enrichment. Second, it neglects gas migration relative to the cluster, although real gas distributions are far more complex than the assumed homogeneous one~\citep{PascaleDaiMcKee_2023ApJ...957...77P}. Third, it ignores the inhomogeneous nature of chemical feedback, where gas around individual stars can be enriched to varying degrees. Finally, the outflow prescription excludes supernova explosions and their feedback, which can inject significant energy after the starburst and rapidly deplete cluster gas. Addressing these limitations requires radiation-hydrodynamic simulations that capture the spatial and temporal complexities of feedback and gas dynamics, offering a more realistic view of chemical enrichment during cluster assembly. We present such simulations below.

\section{Simulations}
\label{sec:sim}

We run magnetohydrodynamical (MHD) simulations of star formation with the code GIZMO~\citep{Hopkins_2015MNRAS.450...53H} in its meshless finite-mass (MFM) mode, which solves the MHD equations in \citet{HopkinsRaives_2016MNRAS.455...51H}. Star formation is modeled with FIRE-3~\citep{HopkinsKeresOnorbe_2014MNRAS.445..581H,HopkinsWetzelKeres_2018MNRAS.480..800H,HopkinsWetzelWheeler_2023MNRAS.519.3154H}, incorporating additional physics like $N$-body gravity, radiation transfer, photo-heating, photo-ionization, photon momentum, mechanical outflows, etc.

The underlying IMF of FIRE-3 does not cover the VMS regime, so we implemented modifications to the standard FIRE physics to model the effects of VMSs~\citep[as in][]{Shi_inprep}. We first quickly recap this framework.

\subsection{Hybrid FIRE and VMS method} \label{sec:sim:hybrid}

We employ a modified version of the FIRE star formation model that, in addition to converting gas cells into ``stars'' representing single stellar populations (SSPs), also forms individually resolved massive stars whose evolution is followed in detail. Thus, the simulation contains two types of ``stars'': FIRE SSP particles representing less-massive stars below a cutoff mass $m_{\rm cut}$, and individual massive stars above it. A detailed description is given in \citet{Shi_inprep}.

This implementation has two main features. The first is the inclusion of a very massive star (VMS) population. In the FIRE-2~\citep{HopkinsWetzelKeres_2018MNRAS.480..800H} and FIRE-3~\citep{HopkinsWetzelWheeler_2023MNRAS.519.3154H} models, stellar evolution and population synthesis adopt an IMF cutoff of $100\,M_\odot$, excluding more massive stars. Our hybrid method extends this to higher cutoffs (e.g., $300\,M_\odot$), distributing newly formed stellar mass between traditional FIRE-like SSP particles and individual VMS particles above a chosen cutoff $m_{\rm cut}$, which are evolved in detail.

The second feature enables simulations of massive GMCs ($\sim10^8\,M_\odot$) with improved efficiency while still resolving individual VMSs. Modules such as STARFORGE~\citep{GrudicGuszejnovHopkins_2021MNRAS.506.2199G,GuszejnovGrudicHopkins_2021MNRAS.502.3646G} model individual stars self-consistently but require extremely high resolution ($\sim0.01\,M_\odot$) and are computationally costly for such large GMCs. By contrast, FIRE simulations are faster at lower resolution ($\sim100$–$1000\,M_\odot$) but lack star-by-star physics. The hybrid method bridges this gap, following the evolution and feedback of individual VMSs analogously to FIRE SSP particles~\citep{HopkinsWetzelKeres_2018MNRAS.477.1578H,HopkinsGrudicWetzel_2020MNRAS.491.3702H}.
In practical simulations, we find that the additional cost to simulate VMSs is insignificant unless an enormous number of VMS particles are created, which causes memory load-balance issues.

In summary, the hybrid method is suitable for studying star formation problems where VMSs play an important role, e.g., the formation and growth of massive remnant BHs of VMSs~\citep{Shi_inprep}. As proved in \S~\ref{sec:toy_model} and Fig.~\ref{fig:toy_model_chemical}, VMSs dominate the chemical feedback by stellar winds; it is then proper to apply the method to simulate this process. 

\begin{figure}
    \centering
    \includegraphics[width=\linewidth]{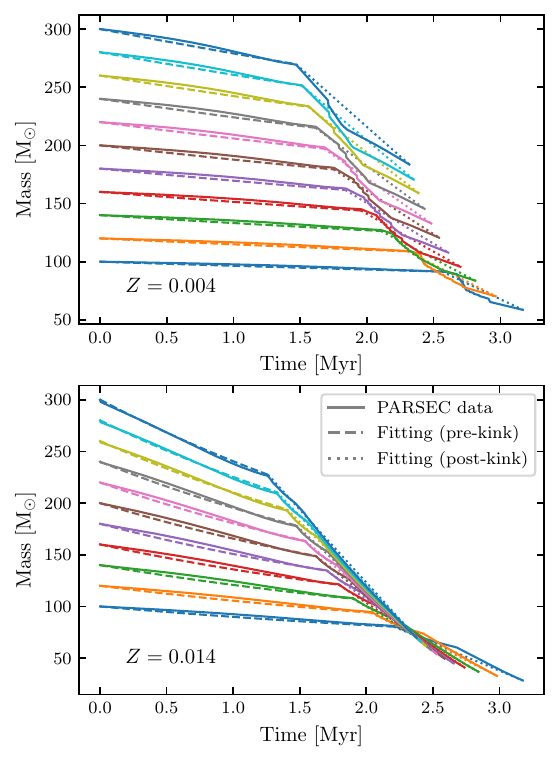}
    \caption{Mass evolution of VMSs compiled from the PARSEC results~\citep{CostaShepherdBressan_2025A&A...694A.193C}. Here we present the evolution tracks (\emph{solid}) for stars with ZAMS mass between $100\,M_\odot$ and $300\,M_\odot$ and $Z=0.004, 0.014$. We also show fittings of these evolution tracks before (following Eq.~\ref{equ:mass_loss_fitting}; \emph{dashed}) and after (linear relation; \emph{dotted}) the ``kink.''
    }
    \label{fig:parsec_mass_loss}
\end{figure}

\subsection{Evolution of VMSs}

We update the VMS mass-evolution prescription of \citet{Shi_inprep} using a new model compiled directly from the PARSEC data. textcolor{black}{Owing to the bistability jump~\citep{VinkdeKoterLamers_1999A&A...350..181V}, the evolution first proceeds through an early hydrogen-burning phase with a low mass-loss rate, followed by a phase with a higher rate, producing a distinct ``kink'' in the mass–time relation (Fig.~\ref{fig:parsec_mass_loss})}. The surface abundances $X_i$ are also different in these phases. The mass-loss rate before $t_{\rm kink}$ is well fitted with
\begin{align}
    \log \dot{m} & = -13.3 + 6.356 \log (m) + 0.9024\log(Z/0.02) \nonumber \\
    & -1.055 \log (m)^2. \label{equ:mass_loss_fitting}
\end{align}
Here $\dot m$ is in the unit of $\rm M_\odot/yr$ and $m$ is in $M_\odot$. In the simulation, for each star with initial mass $m_0$ and metallicity $Z$, we first evolve the stellar mass with the fitting formula. Then we assume a constant mass loss rate after $t_{\rm kink}$ until the star reaches the pre-supernova mass (which is still a good approximation with actual evolutionary tracks, see Fig.~\ref{fig:parsec_mass_loss}). Quantities like $t_{\rm kink}$, stellar lifetime, pre-supernova mass, and remnant mass are functions of both $m_0$ and $Z$, which are obtained as interpolations of tables extracted from the PARSEC dataset.

Fig.~\ref{fig:parsec_mass_loss} shows the mass evolution tracks of stars compiled from the PARSEC data, as well as the fitting relations we used in the simulation, which show good agreement (especially when $t<t_{\rm kink}$). Moreover, the total mass loss in wind for each star is the same as that in PARSEC.

The PARSEC evolution tracks are in agreement with other studies of VMS evolution, that they can lose significant mass in winds before the end of their lifetime \citet{Vink_2018A&A...615A.119V,Vink_2023A&A...679L...9V,Vink_2024arXiv241018980V}. Following the improved mass-loss recipe~\citep{VinkMuijresAnthonisse_2011A&A...531A.132V}, \citet{Vink_2018A&A...615A.119V} found that the terminal wind velocity of VMSs is $\lesssim 500\,\rm km/s$, significantly slower than extrapolations from the $<100\,M_\odot$ massive stars~\citep{LeithererRobertDrissen_1992ApJ...401..596L,LamersSnowLindholm_1995ApJ...455..269L}. Unlike the mass loss rate, the wind velocity is not tightly correlated with the stellar mass. In the simulation, we model the wind velocity $v_{\rm wind}$ with a fitted linear relation based on \citet{Vink_2018A&A...615A.119V}: 
\begin{equation}
    v_{\rm wind}\,[{\rm km/s}] = 573 - 0.326\, m\,[{\rm M_\odot}].
\end{equation}
Compared with predictions in \citet{Vink_2018A&A...615A.119V}, the scatter is $\lesssim 50\,\rm km/s$ in the mass regime of our interest (i.e., $100-300\,M_\odot$).

Like the mass evolution, we model the main-sequence chemical feedback in two stages separated by $t_{\rm kink}$. Following the FIRE convention, we simulate the abundance of 11 elements (H, He, C, N, O, Ne, Mg, Si, S, Ca, Fe). The mass fractions of each element in these three stages are compiled from the PARSEC dataset (as a function of $m_0$ and $Z$), which are constants during each stage, but the total ejected mass of each element in each stage matches the PARSEC data.

In the supernova stage after the main sequence, we distribute the mass in supernova winds into 11 elements following an interpolation table compiled from PARSEC (as a function of $m_0$ and $Z$). We also note that not all VMSs end up with a supernova explosion ~\citep{HegerFryerWoosley_2003ApJ...591..288H,CostaShepherdBressan_2025A&A...694A.193C} and this leads to some modifications to the feedback from FIRE SSP particles (\S~\ref{sec:sim:fire}).

\subsection{Changes to the FIRE-3 physics}
\label{sec:sim:fire}

\begin{figure}
    \centering
    \includegraphics[width=\linewidth]{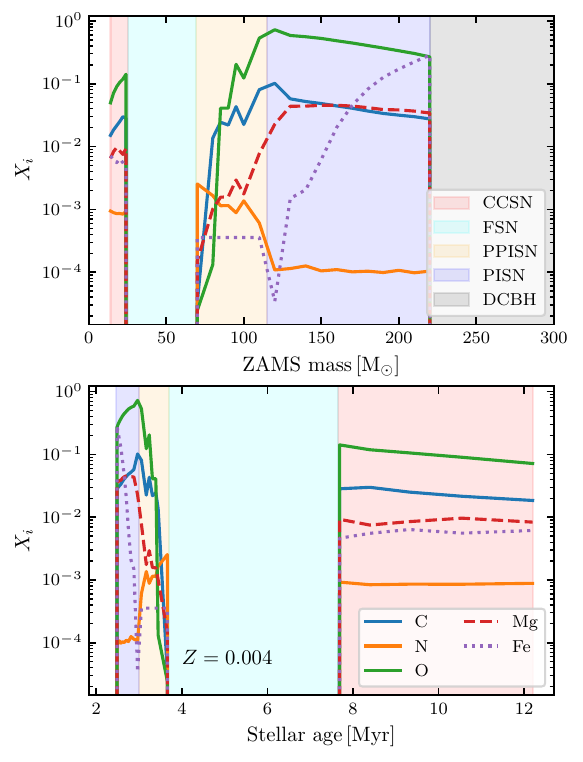}
    \caption{The SNe chemical yield of massive stars and VMSs compiled from the PARSEC data set, for stars with initial metallicity $Z=0.004$. \emph{Top panel} - the SNe yield as a function of the zero-age main-sequence (ZAMS) stellar mass for selected elements (C, N, O, Mg, Fe); note that the fate of a star of a given metallicity is dependent on its ZAMS mass (color shaded regions), including core-collapse SN (CCSN), failed SN (FSN), pulsational pair-instability SN (PPISN), pair-instability SN (PISN), and direct-collapse black hole (DCBH). \emph{Bottom panel} - the SNe yield as a function of stellar age.
    }
    \label{fig:parsec_yield}
\end{figure}

We introduce two changes to the standard FIRE-3 physics. The first concerns the feedback strength of FIRE SSPs, which depends on the IMF cutoff $m_{\rm max}$ through the light-to-mass (L/M) ratio $\xi$, defined as $\xi (m_{\rm max}) = \int_{m_{\rm min}}^{m_{\rm max}} f(m)\, L(m)\,\dd m / \int_{m_{\rm min}}^{m_{\rm max}} f(m)\,m\,\dd m$. The standard FIRE-3 assumes a Kroupa IMF extending to $m_{\rm max}=100\,M_\odot$, yielding $\xi\simeq800\,L_\odot/M_\odot$ for $t\lesssim1\,\rm Myr$~\citep{HopkinsWetzelWheeler_2023MNRAS.519.3154H}. textcolor{black}{With the FIRE+VMS treatment, the underlying IMF of FIRE SSPs extends to $m_{\rm cut}$, so we rescale the luminosity of each FIRE SSP particle by $\xi(m_{\rm cut})/\xi(100\,M_\odot)$ \citep[also see][]{GuszejnovHopkinsMa_2017MNRAS.472.2107G,SuHopkinsHayward_2018MNRAS.480.1666S}.} Individual VMS luminosities are computed from a luminosity–mass relation consistent with FIRE-3 physics~\citep{HopkinsWetzelWheeler_2023MNRAS.519.3154H}.

Moreover, we modify the supernovae (SNe) feedback model of FIRE-3, which includes the impact from both core-collapse SNe and SNe Ia of each FIRE SSP particle 
\citep{HopkinsWetzelWheeler_2023MNRAS.519.3154H}. The early stage of the SNe feedback in FIRE-3 is dominated by core-collapse SNe once the age of the stellar population is older than $3.7\,\rm Myr$. However, from more detailed stellar evolution simulations with PARSEC in \citet{CostaShepherdBressan_2025A&A...694A.193C}, massive stars of $\sim 30$--$75\,M_\odot$ (depending on metallicity) may only terminate as ``failed SNe'' that inject little mass and energy into the interstellar medium due to their strong self-gravity, as indicated by theory~\citep{HegerFryerWoosley_2003ApJ...591..288H} and observations~\citep[e.g., N6946-BH1,][]{GerkeKochanekStanek_2015MNRAS.450.3289G,BeasorHosseinzadehSmith_2024ApJ...964..171B}. Instead, (``successful'') core-collapse SNe can only appear after $t_{\rm CC} \sim 7$--$7.8\,\rm Myr$ (depending on metallicity), corresponding to the lifetime of $m\sim 30\,M_\odot$ stars. We find $t_{\rm CC}$ can be well fitted with
\begin{align}
    t_{\rm CC} = \begin{cases}
        8.1 - 0.1\,\log (Z), & Z\leq 10^{-3}; \\
        4.0 - 2.8\,\log (Z) -0.5\,\log(Z)^2, & Z > 10^{-3}.
    \end{cases}
\end{align}
Here, the unit of $t_{\rm CC}$ is Myr. Consequentially, we set $m_{\rm cut}=75\,M_\odot$ and only turn on the FIRE-3 SNe feedback if $t>t_{\rm CC}$ so the effects of ``failed SNe'' are modeled, so both the kinetic and chemical feedback are not overestimated.

Figure~\ref{fig:parsec_yield} illustrates that $Z=0.004$ stars produce different remnants depending on their ZAMS masses (\emph{upper panel}). Within our IMF range, the most massive stars ($\gtrsim220\,M_\odot$) collapse directly into black holes (DCBHs; gray), while VMSs of $70$--$220\,M_\odot$ end as pulsational pair-instability SNe (PPISN; orange) or pair-instability SNe (PISN; blue), yielding significant mass loss and chemical enrichment. Stars of $25$--$70\,M_\odot$ become ``failed SNe'' (FSN; cyan), and those below $25\,M_\odot$ explode as core-collapse SNe (CCSN; red) or lower-energy events such as SNe~Ia (not shown). The \emph{bottom panel} shows SN yields versus stellar age, illustrating the feedback model adopted in our simulations: DCBH and PPISN progenitors ($m_{\rm ZAMS}\gtrsim m_{\rm cut}=75\,M_\odot$) are modeled as individual VMSs evolved until $\sim3.7\,\rm Myr$; the FSN phase ($\sim3.7$--$7.8\,\rm Myr$) contributes no chemical feedback; and the subsequent CCSN phase follows the FIRE-3 SSP treatment~\citep{HopkinsWetzelWheeler_2023MNRAS.519.3154H}.

\begin{deluxetable*}{lcccl}
\tablecaption{Simulation runs performed in this study.}
\label{tab}
\tablenum{1}
\tablehead{\colhead{Simulation group} & \colhead{$M_{\rm cloud}\,[M_\odot]$} & \colhead{$R_{\rm cloud}\,[\rm pc]$} & \colhead{$Z\,[Z_\odot]$} & \colhead{Star formation framework} } 
\startdata
    Fiducial simulations (\S~\ref{sec:res:fiducial}) & $3\times 10^7$ &60, 70, 80& 0.15, 0.2, 0.25 & FIRE+VMS \\
    GMC surface density tests (\S~\ref{sec:res:surface_density}) & $10^7, 10^8$ &60, 70, 80& 0.2 & FIRE+VMS \\
    FIRE-only simulations (\S~\ref{sec:res:fire}) & $3\times 10^7$ &60, 70, 80& 0.2 & FIRE (unaltered, or with FSN) \\
    VMS feedback tests (\S~\ref{sec:res:feedback_tests}) & $3\times 10^7$ &60, 70, 80& 0.2 & FIRE+VMS (disabling feedback) \\
\enddata
\end{deluxetable*}

\begin{figure*}
    \centering
    \includegraphics[width=\linewidth]{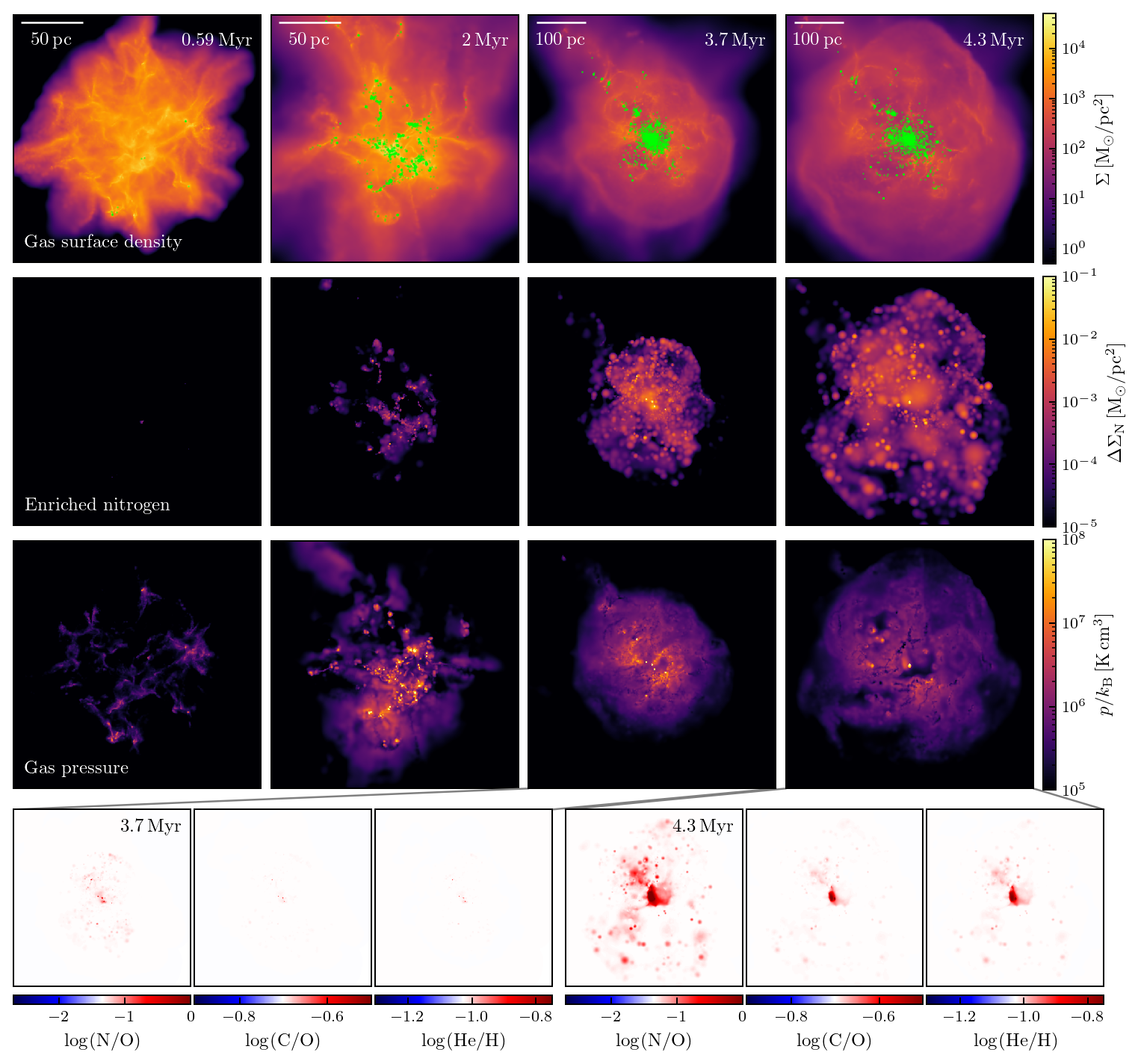}
    \caption{Visualization of the simulation \path{M3e7_R70_Z0.2}. Here we show the snapshots at different simulation times (0.59, 2, 3.7, 4.3 Myr) in the four columns. From top to bottom, in each row is: 1) the gas surface density and locations of VMSs (green dots); 2) the mass surface density of nitrogen that is enriched by stars (and SNe); 3) the line-averaged gas pressure near the mid-plane; 4) the chemical abundance ratios (N/O, C/O, and He/H) near the mid-plane at $t=3.7\,\rm Myr$ and $4.3\,\rm Myr$ (as indicated by the guiding lines). The white color in the color bar corresponds to values of the natal GMC, and there is no significant enrichment before 3.7\,Myr. 
    }
    \label{fig:simulation_visualization}
\end{figure*}

\subsection{Simulation suite}
\label{sec:sim:suite}

The star formation physics in this study follows FIRE-3~\citep{HopkinsWetzelWheeler_2023MNRAS.519.3154H}, which evaluates the GMC star formation rate while modeling the unresolved IMF with the hybrid method described in \S~\ref{sec:sim:hybrid}. A gas cell forms a FIRE SSP particle only if (1)~its density exceeds $1000\,\rm cm^{-3}$, (2)~it is self-gravitating against turbulence, thermal, and magnetic support, (3)~it exhibits convergent inflow ($\nabla\cdot\mathbf{v}<0$), and (4)~its Jeans mass is smaller than the cell mass~\citep{GrudicGuszejnovHopkins_2021MNRAS.506.2199G}. The star formation rate is further weighted by the cell’s molecular fraction~\citep{HopkinsWetzelKeres_2018MNRAS.480..800H}. Once an SSP forms, Poisson sampling determines the number of VMSs, which are spawned as individual particles~\citep[see][]{Shi_inprep}.

Radiative (UV, optical, IR) and kinetic (winds and SNe) feedback follow \citet{HopkinsWetzelKeres_2018MNRAS.477.1578H,HopkinsWetzelKeres_2018MNRAS.480..800H}. Radiation transfer is treated as in \citet{HopkinsQuataertMurray_2012MNRAS.421.3488H}, where the luminosity $L_\nu$ in each band is locally attenuated by gas and dust before propagating into the ISM, producing momentum coupling and radiative heating/cooling. In the UV, where massive stars dominate, the flux-weighted opacity is $\kappa_\nu = 1800\,(Z/Z_\odot)\,{\rm cm}^2\,{\rm g}^{-1}$, assuming dust abundance scales with metallicity. VMS luminosities in each band are computed assuming blackbody spectra, following \citet{GrudicGuszejnovHopkins_2021MNRAS.506.2199G}.

Metals are treated as passive scalars, evolved via the advection–diffusion equation with stellar yield source terms, following \citet{HopkinsWetzelKeres_2018MNRAS.480..800H} and \citet{RennehanBabulHopkins_2019MNRAS.483.3810R}, and verified by \citet{ColbrookMaHopkins_2017MNRAS.467.2421C}. In addition to resolved mixing through the continuity equation, unresolved small-scale transport is modeled with the Smagorinsky turbulent diffusion prescription~\citep{Smagorinsky_1963MWRv...91...99S,EscalaWetzelKirby_2018MNRAS.474.2194E}.  

The initial conditions follow previous setups~\citep{GrudicHopkinsFaucher-Giguere_2018MNRAS.475.3511G,Shi_inprep}: a spherical, uniform, non-rotating GMC with turbulent velocities containing 50\% solenoidal modes and total kinetic energy equal to the gravitational binding energy, $\alpha_{\rm vir}\!\equiv\!2T/|U|\!=\!2$. Although observations show a broad $\alpha_{\rm vir}$ distribution~\citep[e.g.,][]{SunLeroySchinnerer_2020ApJ...901L...8S}, this value is typical for massive GMCs~\citep{ChevanceKrumholzMcLeod_2023ASPC..534....1C}.
The initial magnetic field is uniform, with magnetic energy set to 1\% of the gravitational binding energy. The initially uniform density and field quickly become turbulent owing to the imposed velocity field. Elemental abundances follow the empirical relations in Eqs.~\eqref{equ:no_fitting} and \eqref{equ:co_fitting}.

Each simulation is labeled \verb|M%g_R%g_Z%g|, denoting the GMC mass $M_{\rm cloud}$ ($M_\odot$), radius $R_{\rm cloud}$ (pc), and metallicity $Z$ ($Z_\odot$); e.g., \verb|M3e7_R70_Z0.2| corresponds to $M_{\rm cloud}=3\times10^7\,M_\odot$, $R_{\rm cloud}=70\,{\rm pc}$, and $Z=0.2\,Z_\odot$. All GMCs start with $64^3$ equal-mass gas cells and evolve for 10\,Myr. When VMSs are included, we adopt a Kroupa IMF extending to $m_{\rm max}=300\,M_\odot$, with stars above $m_{\rm cut}=75\,M_\odot$ treated individually and less massive ones modeled as FIRE SSPs. We run four simulation groups for different goals (Table~\ref{tab}).

\begin{enumerate}
    \item \emph{Fiducial simulations}. We use the FIRE+VMS module explained earlier to perform the fiducial simulations. Based on properties of the LyC leaking cluster~\citep{PascaleDaiMcKee_2023ApJ...957...77P}, our fiducial simulations have fixed $M_{\rm cloud} = 3\times 10^7\,M_\odot$ but vary $R_{\rm cloud}$ ($60, 70, 80\,\rm pc$) and $Z/Z_\odot$ (0.2, 0.15, 0.25).
As will be explained in \S~\ref{sec:res}, these simulations can (roughly) reproduce the observations. 

    \item \emph{GMC surface density tests}. This variation keeps the fiducial setup but initializes with different $M_{\rm cloud}\, (10^7, 10^8\,\rm M_\odot)$, while keeping the same $R_{\rm cloud}$ ($60, 70, 80\,\rm pc$) and $Z/Z_\odot=0.2$. These simulations explore the impact of the GMC initial mass (and surface density), to accommodate the order-of-unity uncertainty in the star cluster's mass due to the magnification factor of gravitational lensing~\citep{PascaleDaiMcKee_2023ApJ...957...77P}.

    \item \emph{FIRE-only simulations}. These simulations remove explicitly-modeled VMSs and utilize FIRE-3 physics (where the underlying IMF only extends to $100\,M_\odot$) to highlight the effect of VMSs in shaping the star formation, especially the chemical evolution. They initialize with $M_{\rm cloud} = 3\times 10^7\,M_\odot$, $R_{\rm cloud} = 60, 70, 80\,\rm pc$, and $Z=0.2\,Z_\odot$. In particular, since we removed ``failed SNe'' from FIRE-3 when implementing the hybrid FIRE+VMS model (\S~\ref{sec:sim:fire}), we also run such a version of simulations in contrast to the standard FIRE-3 physics.

    \item \emph{VMS feedback tests}. These simulations aim to explore how VMSs impact star formation through different forms of feedback (radiation, winds, SNe). They are based on simulations with $M_{\rm cloud} = 3\times 10^7\,M_\odot$, $R_{\rm cloud} = 60, 70, 80\,\rm pc$, and $Z=0.2\,Z_\odot$, but we disable selected forms of feedback by re-normalizing their energy with a very small number. 
\end{enumerate}

\section{Results}
\label{sec:res}

\begin{figure*}
    \centering
    \includegraphics[width=\linewidth]{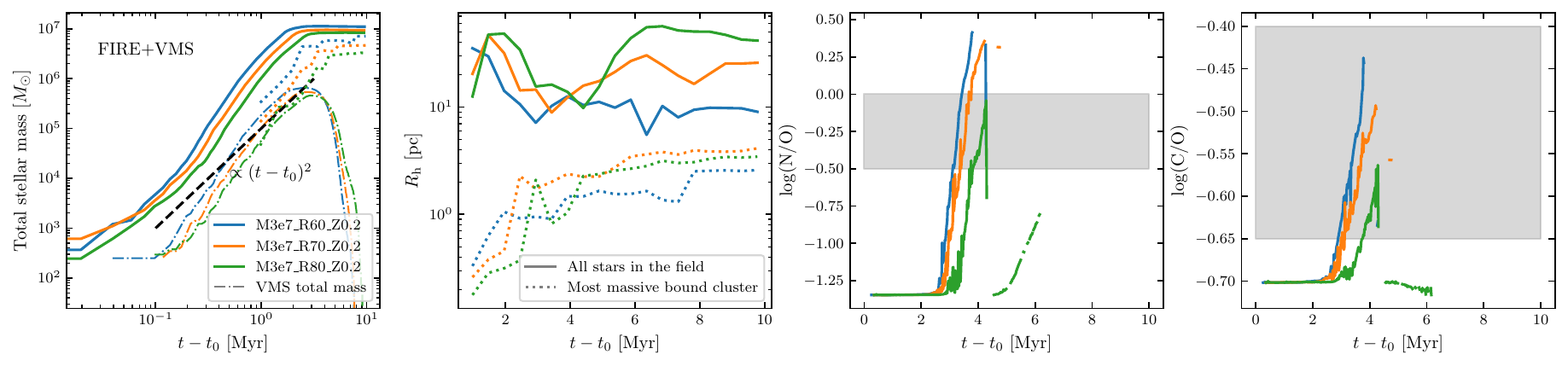}
    \includegraphics[width=\linewidth]{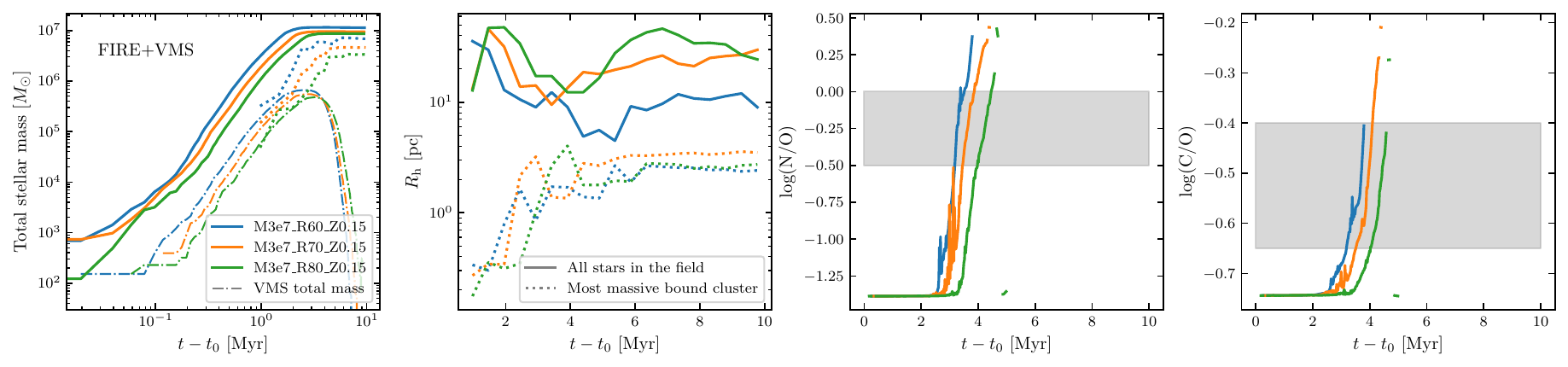}
    \includegraphics[width=\linewidth]{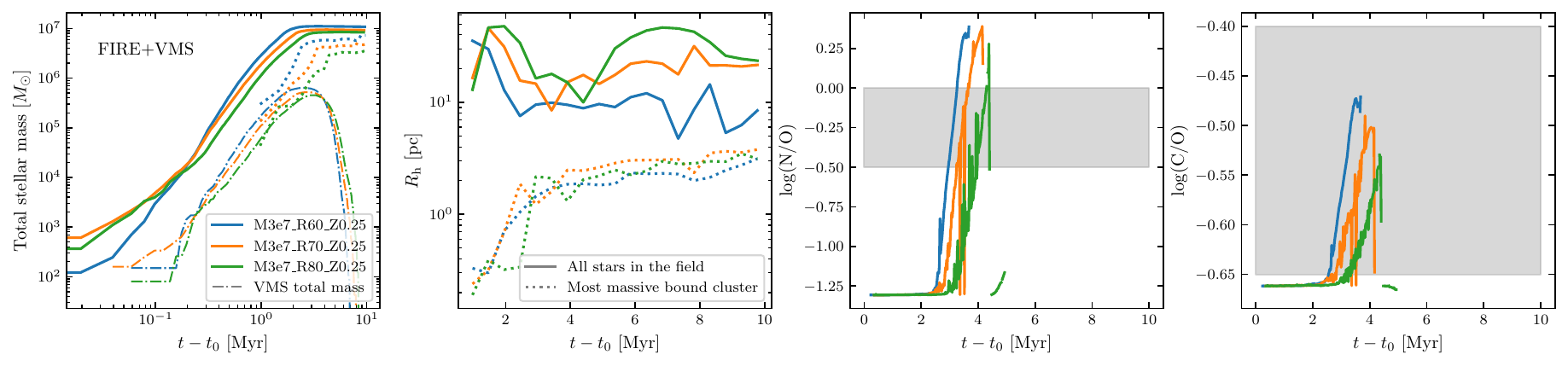}
    \caption{
    Star formation and the chemical abundance of high-pressure ($>10^9\,\rm K\,cm^{-3}$) gas throughout the evolution of our ``fiducial'' group of simulations, as a function of ``time since star formation.'' These simulations initialize with different GMCs (\path{M3e7_R60}, \path{M3e7_R70}, \path{M3e7_R80}) and metallicities: $0.2\,Z_\odot$ (\emph{top row}), $0.15\,Z_\odot$ (\emph{second row}), and $0.25\,Z_\odot$ (\emph{bottom row}). Each row, from left to right, shows the 1) star formation history, in general (\emph{solid}), or of the most massive bound cluster (\emph{dotted}), or of all VMSs (\emph{dot dashed}); 2) UV half-light radius (of all stars in the field; \emph{solid}) and the 3D half-mass radius (of the most massive bound cluster; \emph{dotted}); 3) median $\log(\rm N/O)$ for the high-pressure gas; 4) median $\log(\rm C/O)$ for the high-pressure gas. We also show the rough range of abundance ratios from \citet{PascaleDaiMcKee_2023ApJ...957...77P} as gray shaded regions.
    }
    \label{fig:fiducial_results}
\end{figure*}

\subsection{Fiducial results}
\label{sec:res:fiducial}

We first go through our ``fiducial'' simulations (\S~\ref{sec:sim:suite}).

Figure~\ref{fig:simulation_visualization} visualizes the \path{M3e7_R70_Z0.2} simulation. The top row shows the gas surface density and VMS distribution at several times. By $\sim0.59\,\rm Myr$, the GMC fragments into dense clumps, forming stars (green dots) off-center. At $\sim2\,\rm Myr$, hierarchical sub-clusters develop during further collapse, and by $\sim3.7\,\rm Myr$, they merge near the cloud center. Strong stellar feedback then expels the surrounding gas, driving large-scale expansion by $\sim4.3\,\rm Myr$.

The second row shows the surface density of enriched nitrogen, $\Delta\Sigma_{\rm N}$. Each gas cell of mass $m_j$ and initial abundance $X_{i,j}$ (for element $i$) evolves to $m_j'$ and $X_{i,j}'$ after star formation and feedback. The enriched mass is defined as $\Delta m_{i,j}=m_j'X_{i,j}'-m_jX_{i,j}$, from which we compute $\Delta\Sigma_{\rm N}$. As stars form, nitrogen enrichment becomes visible: at $\sim2\,\rm Myr$, $\Delta\Sigma_{\rm N}$ peaks at star-forming regions (contrast the nearly blank map at $\sim0.6\,\rm Myr$); by $\sim3.7\,\rm Myr$, feedback enhances central enrichment, and by $\sim4.3\,\rm Myr$, the enriched gas is driven outward.

The third row shows the line-of-sight averaged gas pressure, i.e., $\bar p = \int_{-l/2}^{l/2} p(\mathbf{x}) \dd z / l$, where $l$ equals the side length of the field of view for each panel. Even averaged along the line of sight, the pressure reaches as high as $\sim 10^8\,\rm K\, cm^{-3}$ at $\sim 2\,\rm Myr$ in the vicinity of stars (like the enriched nitrogen). At $\sim 4.3\,\rm Myr$, there is a pressure peak at the center of the cluster, suggesting a compact ($< 1\,$pc) gas blob pressurized by stellar feedback. 

The bottom row shows the abundance ratios N/O, C/O, and He/H. They remain close to the initial GMC values except at late times ($\sim4.3\,\rm Myr$), when enrichment occurs rapidly within $\lesssim0.6\,\rm Myr$, given the absence of signals at $\sim3.7\,\rm Myr$. Notably, $\log(\mathrm{N/O})$ rises from $\sim-1.2$ to $\sim0$, consistent with the enrichment level observed in the high-pressure gas of the Sunburst LyC cluster~\citep{PascaleDaiMcKee_2023ApJ...957...77P}.

Although the second row of Fig.~\ref{fig:simulation_visualization} shows nitrogen widely dispersed throughout the cloud, the bottom row indicates that strongly elevated N/O (and similarly enhanced C/O and He/H) appears only within $\sim10$\,pc of the super star cluster. 
Such extreme N/O values arise in only a small fraction of the enriched gas, a point explored further below.

\begin{figure*}
    \centering
    \includegraphics[width=\linewidth]{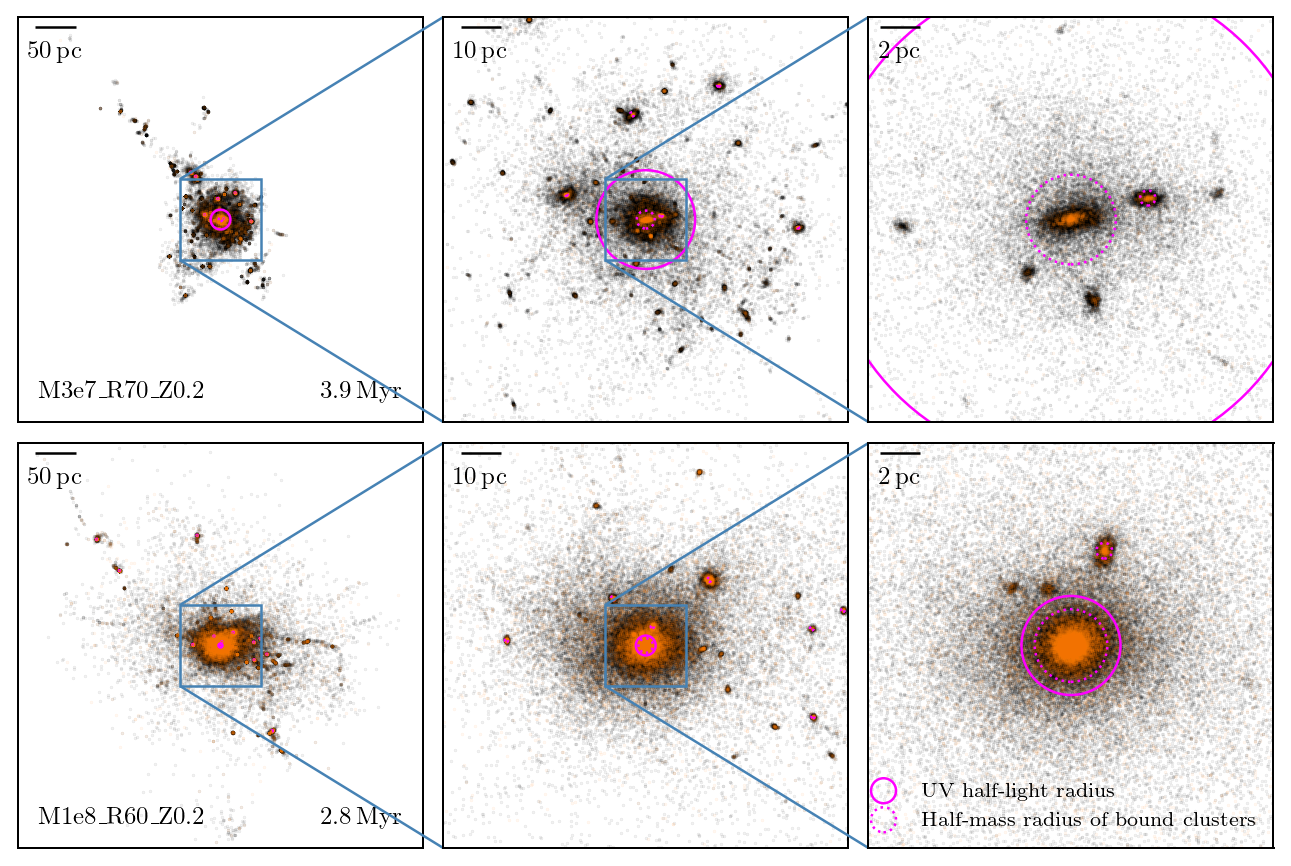}
    \caption{Distribution of stars in two simulations \path{M3e7_R70_Z0.2} (\emph{top}) and \path{M1e8_R60_Z0.2} (\emph{bottom}) at the time when $\log(\rm N/O)>-0.25$ for the high-pressure ($>10^9\,\rm K\,cm^{-3}$) gas. Here we show the scatter plot of two types of ``stars'': FIRE SSPs (\emph{black dots}) and individual VMSs (\emph{orange dots}). From left to right, we zoom in towards the most massive sub-cluster. The UV half-light radius of this whole distribution of stars, as well as the half-mass radii of identified bound clusters, are also marked. 
    }
    \label{fig:star_cluster_scattering}
\end{figure*}

\subsubsection{Super star cluster formation}

The left two columns of Fig.~\ref{fig:fiducial_results} characterize the evolution of the simulated star cluster. For each simulation, we identify the onset time of star formation $t_0$, which is used to define the cluster age $t-t_0$. The exact time $t_0$ is different in each simulation but is typically around $\sim 0.1\,\rm Myr$, suggesting that star formation is triggered soon after GMC collapse. The mass of newly-formed stars initially increases as $M_\star \propto (t-t_0)^2$~\citep[][]{LiKleinMcKee_2018MNRAS.473.4220L}, but the growth becomes steeper later after $t-t_0\sim 0.3\,\rm Myr$. Moreover, star formation culminates at $\sim 1\,\rm Myr$, when half of the stellar mass ($\sim 5\times 10^6\,M_\odot$) forms. Star formation is completed at $\sim 2-3\,\rm Myr$, roughly within two global free-fall times of the natal cloud.

The cumulative stellar mass formed in our fiducial simulations (\path{M3e7_R60}, \path{M3e7_R70}, \path{M3e7_R80}) reaches $\sim 10^7\,M_\odot$, which is similar to estimated stellar mass of the Sunburst Arc LyC cluster. The cumulative stellar mass is slightly lower for GMCs with larger initial radii, but this is not a significant effect in the parameter space explored here. There is also little variation within the metallicity range $0.15$--$0.25\,Z_\odot$ we explore.

Fig.~\ref{fig:star_cluster_scattering} shows that star formation first leads to multiple gravitationally bound sub-clusters instead of a major cluster. In the first column of Fig.~\ref{fig:fiducial_results}, we also show the mass growth of the central cluster (\emph{dotted lines}), which reaches $>3\times 10^6\,M_\odot$ and accounts for an order-unity fraction of the cumulative stellar mass. We note the trend that more compact GMCs give rise to a more massive central cluster that accounts for a larger fraction of the cumulative stellar mass. 

We also plot the time evolution of the total mass of VMSs in the same panels. The total mass of VMSs first grows with time with a slope similar to the total stellar mass (but is smaller by a factor of $\sim 0.1$ as expected from the IMF), and then reaches the peak of $\sim 5\times 10^5\,M_\odot$ at $\sim 3\,$Myr, before it declines as a result of VMS mass loss and death. At $t-t_0\sim 4\,\rm Myr$, the VMS total mass declines to $\sim 3\times 10^5\,\rm M_\odot$, which implies $\sim 2000$ VMSs by assuming the IMF-averaged VMS mass of $\sim 150\,M_\odot$. \citet{MestricVanzellaUpadhyaya_2023A&A...673A..50M} find that the presence of $\sim 400$ VMSs is needed to explain the VMS stellar wind features in the Sunburst LyC cluster, significantly less than the cumulative number of VMSs expected from the default IMF and formed in our simulation. This may or may not reflect a genuine effect on the IMF, and we caution about the uncertainty in predicting the number of surviving VMSs present in the LyC cluster, due to the uncertainty in VMS evolution and cluster age.

The second column of Fig.~\ref{fig:fiducial_results} shows the time evolution of the UV half-light radius for all the stars in the field.
The radius is measured relative to the center of the most massive cluster (which also marks the luminosity peak), defined by summing the UV luminosities of all star particles while neglecting ISM radiative transfer. For each star, the bolometric luminosity $L_{\rm bol}(m)$ follows the luminosity–mass relation of \citet{HopkinsWetzelKeres_2018MNRAS.480..800H}, and the stellar radius $R_\star(m)$ follows \citet{ToutPolsEggleton_1996MNRAS.281..257T}. The effective temperature is $T_{\rm eff} \approx 5770\,{\rm K}\,[(L_{\rm bol}/R_\star^2)/(L_\odot/R_\odot^2)]^{1/4}$. The UV luminosity is $L_{\rm UV} = f_{\rm UV} L_{\rm bol}$, where $f_{\rm UV}$ is the fraction of black-body emission in the FIRE UV band ($3.444 < h\nu/{\rm eV} < 8$). Massive stars dominate the UV output: $L_{\rm UV}/m \approx 0.08\,L_\odot/M_\odot$ at $1\,M_\odot$, increasing to $L_{\rm UV}/m \approx 7250\,L_\odot/M_\odot$ at $300\,M_\odot$. We also show the half-mass radius of the most massive cluster.

The UV half-light radius is typically in the range $\sim 10-50\,\rm pc$ throughout the cluster formation process, but can be as small as $\lesssim 10\,\rm pc$ at a cluster age $\sim 2$--$4\,\rm Myr$, which is compatible with the Sunburst LyC cluster. On the other hand, the half-mass radius of the central cluster monotonically increases with time. At $t-t_0\sim 4\,\rm Myr$, the increase saturates at $\sim 2$--$3\,\rm pc$, which is still significantly smaller than the UV half-light radius.

These results suggest that the Sunburst LyC cluster may consist of many low-mass satellite sub-clusters surrounding a central massive compact cluster. This could explain the resolved outskirts of UV emission~\citep{PascaleDaiMcKee_2023ApJ...957...77P, VanzellaCastellanoBergamini_2022A&A...659A...2V}. In our simulations, the satellites originate from the same natal cloud and are coeval with the central cluster.

Fig.~\ref{fig:star_cluster_scattering} also shows the spatial distribution of VMSs, which are concentrated at the center of each sub-cluster, which is expected since dynamical friction is faster for the more massive stars~\citep{Chandrasekhar_1943ApJ....97..255C}. Such an effect is modeled in our simulation with a tree-based algorithm~\citep{MaHopkinsKelley_2023MNRAS.519.5543M}, and aligns with the idea of mass segregation put forth in \citet{MestricVanzellaUpadhyaya_2023A&A...673A..50M}. Moreover, the oblate shape of a subset of sub-clusters in Fig.~\ref{fig:star_cluster_scattering} may suggest significant spins of young star clusters formed in turbulent GMCs.

\begin{figure*}
    \centering
    \includegraphics[width=\linewidth]{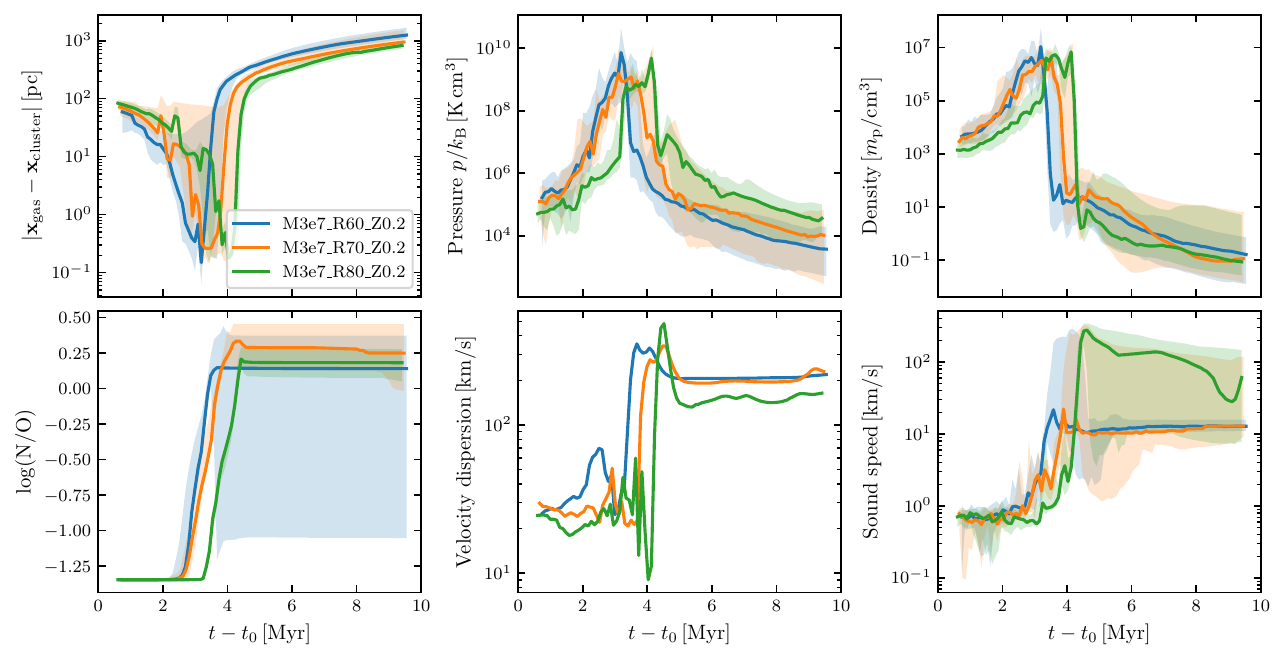}
    \caption{Evolution of the nitrogen-enriched gas packet for simulations with $M_{\rm cloud}=3\times 10^7\,M_\odot$ (i.e., fiducial runs). We show properties of the nitrogen-enriched gas, including 1) distance to the center of the most massive bound cluster, $\mathbf{x}_{\rm center}$; 2) pressure; 3) density; 4) N/O ratio; 5) velocity dispersion; 6) sound speed. Shaded regions represent the $16\%$ and $84\%$ percentiles. We caution that, unlike Fig.~\ref{fig:fiducial_results}, which selects and traces high-pressure ($>10^9\,\rm K\,cm^{-3}$) gas, the gas packet includes the same ensemble of gas cells throughout the simulation.
    }
    \label{fig:high_no}
\end{figure*}

\subsubsection{Chemical enrichment}

The right two columns of Fig.~\ref{fig:fiducial_results} show N/O and C/O ratios in the high-pressure gas, which is selected for gas cells with $p> 10^9 \rm K\,\rm cm^{-3}$~\citep[following][]{PascaleDaiMcKee_2023ApJ...957...77P}, and the curves in these panels represent the median $\log (\rm N/O)$ and $\log (\rm C/O)$ values of these gas cells. Compared with the observed abundance ratios for the Sunburst LyC cluster from \citet{PascaleDaiMcKee_2023ApJ...957...77P} (\emph{gray shaded regions}), we find that the fiducial simulations can reproduce the observed N/O and C/O abundance ratios during the period $2\,{\rm Myr}<t-t_0 \lesssim 4\,\rm Myr$. It is also possible that $\log(\rm N/O)$ can even rise above $0$, before the curves break or disappear at $\sim 4\,\rm Myr$ (which suggests absence of high-pressure gas). As also indicated in Fig.~\ref{fig:simulation_visualization}, during this time, stellar feedback becomes effective and continuously pushes the gas envelope outward, eventually leading to depletion of the high-pressure gas. 

\begin{figure*}
    \centering
    \includegraphics[width=\linewidth]{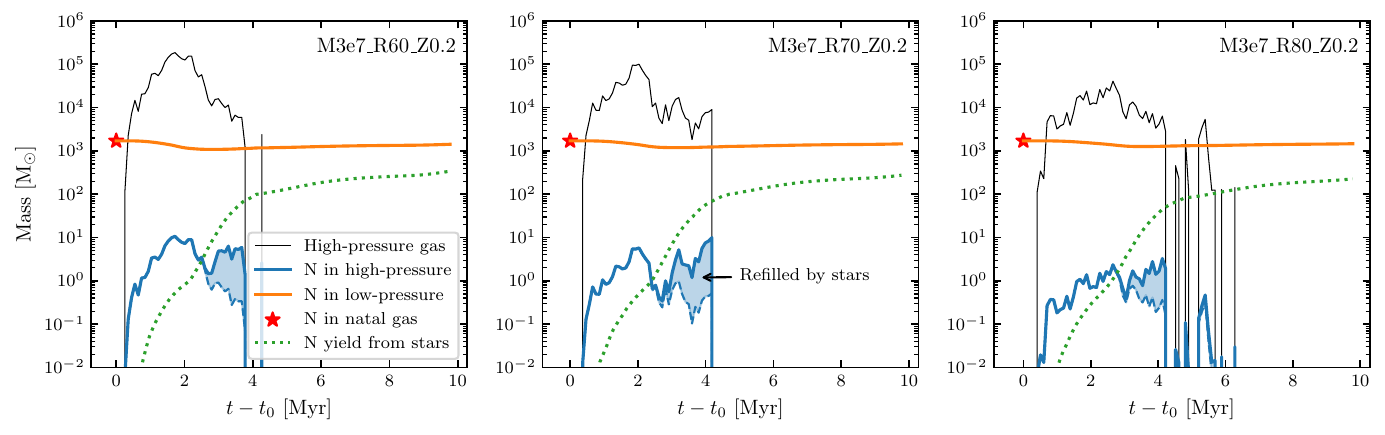}
    \caption{The mass budget of nitrogen for our fiducial simulations. For each simulation, we first show the mass of the high-pressure ($>10^9\,\rm K\,cm^{-3}$) gas (\emph{gray solid}). On top of it are the nitrogen masses in the high-pressure gas and the low-pressure ($<10^9\,\rm K\,cm^{-3}$) gas. In particular, we also show the pristine nitrogen mass in the high-pressure gas (\emph{blue dashed}) to highlight the amount refilled by stellar winds (\emph{blue shaded region}).
    }
    \label{fig:nitrogen_budget}
\end{figure*}

From Fig.~\ref{fig:fiducial_results}, there is also a clear trend that more compact GMCs with a shorter free-fall time tend to exhibit the required high N/O value faster. However, all GMC initial conditions explored in Fig.~\ref{fig:fiducial_results} are in reasonable agreement with observations. By checking simulations with $Z=0.15\,Z_\odot$ and $Z=0.25\,Z_\odot$, we find the results robust in this range of metallicity.

To further investigate the chemical enrichment process, Fig.~\ref{fig:high_no} traces the nitrogen-enriched gas packet, i.e, a fixed ensemble of gas cells in our Lagrangian simulation. These gas cells are selected as high-pressure ($>10^9\,\rm K\,cm^{-3}$) gas at the time when their median $\log(\rm N/O)$ exceeds $-0.25$ (practically, their unique particle IDs are marked to form the gas packet). We note that Fig.~\ref{fig:high_no} shows ``how this gas packet gets nitrogen-enriched,'' while the right two columns of Fig.~6 show N/O and C/O ratios of the high-pressure ($>10^9\,\rm K\,cm^{-3}$) gas cells selected at each time, so the ensemble of gas cells is being dynamically updated.

We find that the nitrogen-enriched gas undergoes two dynamical phases. 
Initially, it sinks toward the center of the massive cluster as the separation $|\mathbf{x}_{\rm gas}-\mathbf{x}_{\rm center}|$ drops to $\sim0.1\,\rm pc$ for $t-t_0\!\sim\!3$–$4\,\rm Myr$, where it mixes with stellar-wind material. This agrees with \citet{Rivera-ThorsenChisholmWelch_2024A&A...690A.269R} that the enriched gas should be condensed inside the cluster and close to the most massive stars.
During this stage, the pressure rises to $\gtrsim10^9\,\rm K\,cm^{-3}$ and the density to $\gtrsim10^6\,m_{\rm p}\,{\rm cm^{-3}}$. 
Enrichment occurs at $\sim3\,\rm Myr$, before SNe, with velocity dispersion $\sim10$–$20\,\rm km\,s^{-1}$—comparable to but below that inferred for the LyC cluster from nebular-line widths ($\sim40\,\rm km\,s^{-1}$)~\citep{VanzellaCastellanoBergamini_2022A&A...659A...2V}, indicating the gas remains gravitationally bound. 
The sound speed remains $\sim1\,\rm km\,s^{-1}$ until $\sim3\,\rm Myr$, then rises to $\sim10\,\rm km\,s^{-1}$, signaling photoionization. 
The evolution of density and pressure thus reflects an isothermal collapse followed by stellar photoheating.

After enrichment, the high-pressure gas is expelled to $\gtrsim100\,\rm pc$, with velocity dispersion increasing to $\sim200\,\rm km\,s^{-1}$, unbinding it from the cluster. 
As external confinement is lost, both pressure and density decline, while the $\sim10\,\rm km\,s^{-1}$ sound speed remains consistent with photoionized gas. 
This phase marks the efficient removal of polluted gas (natal material mixed with stellar winds). 
The nitrogen-enriched gas retains a high N/O ratio but with a large variance, likely due to turbulent mixing of metals.

In summary, the nitrogen-enrichment appears as a temporary phase of star cluster formation from GMC collapse. In terms of mass, the nitrogen-enriched gas is a small subset of the gas complex (also see discussion in \S~\ref{sec:nitrogen_mass_budget}). The enriched gas is typically dense ($\gtrsim 10^5\,\rm cm^{-3}$) and highly-pressurized ($\gtrsim 10^9\,\rm K\,cm^{-3}$), in agreement with what has been observed in the Sunburst LyC cluster and other high-redshift nitrogen-enriched galaxies~\citep{PascaleDai_2024ApJ...976..166P,JiUblerMaiolino_2024MNRAS.535..881J,ToppingStarkSenchyna_2024MNRAS.529.3301T,YanagisawaOuchiWatanabe_2024ApJ...974..266Y}. This gas originates from the natal GMC and settles down at the cluster center following a nearly isothermal gravitational collapse, but is then heated by stellar feedback and eventually ejected from the cluster. 

Recent studies~\citep{RizzutiMatteucciMolaro_2025A&A...697A..96R,McClymontTacchellaSmith_2025arXiv250708787M} suggest that elevated N/O ratio is a temporary phenomenon in the ISM of high-redshift galaxies, and arises from differential winds. Unlike the nitrogen-enrichment process we investigate here, the relevant spatial scale is much larger than the proximity of newborn super star clusters, and the relevant timescale more aligns with the longer evolution of AGB stars rather than with the rapid evolution of VMSs. It is unclear if that picture is compatible with the intriguing observation that the nitrogen-enriched gas is almost always as dense as $\gtrsim 10^5\,\rm cm^{-3}$ in observed galaxies and clumps (references listed above).

\begin{figure*}
    \centering
    \includegraphics[width=\linewidth]{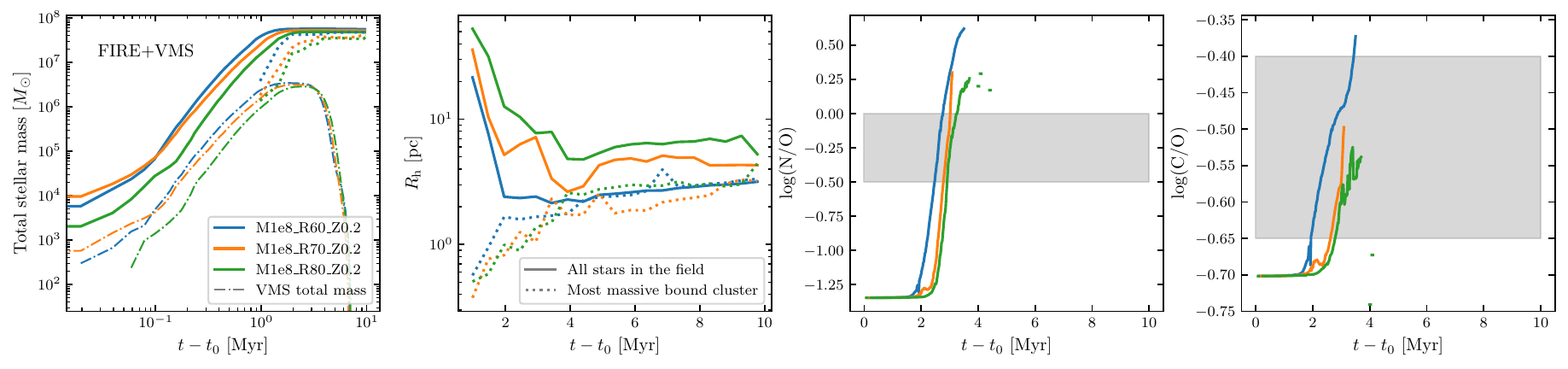}
    \includegraphics[width=\linewidth]{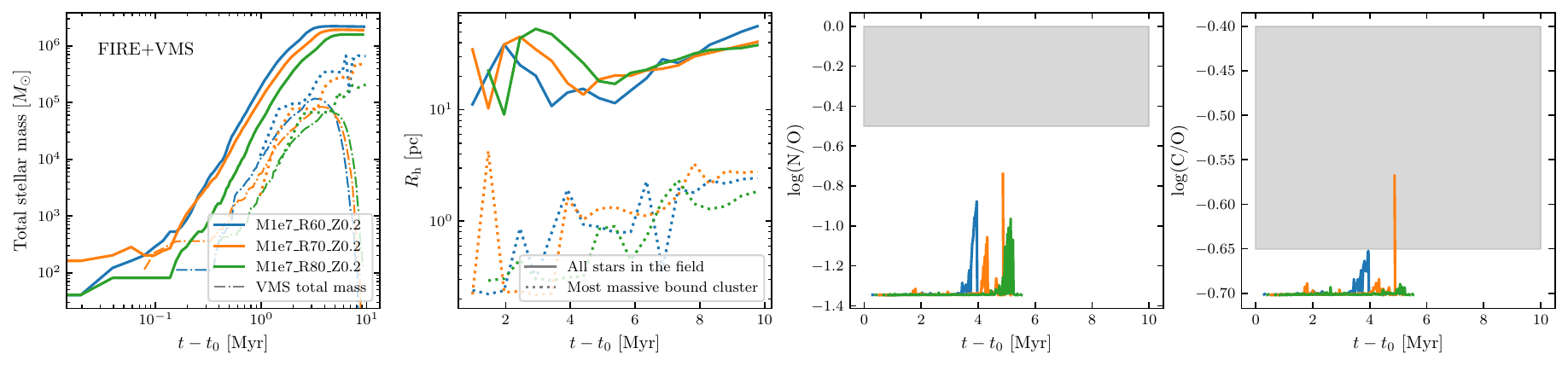}
    \caption{Same as the the setup of fiducial simulations (Fig.~\ref{fig:fiducial_results}), but with different GMC initial conditions of $M_{\rm cloud}=10^8\,M_\odot$ (\emph{top}) and $M_{\rm cloud}=10^7\,M_\odot$ (\emph{bottom}).}
    \label{fig:results_mass_variation}
\end{figure*}

\subsubsection{Nitrogen mass budget}
\label{sec:nitrogen_mass_budget}

As Fig.~\ref{fig:simulation_visualization} shows, a large amount of nitrogen is produced at $\sim 4\,\rm Myr$ after the onset of simulation, such that the surface density of enriched nitrogen reaches $\Delta \Sigma_{\rm N} \sim 10^{-2}\,M_\odot\,{\rm pc}^{-2}$. Below, we study the evolution of the nitrogen mass to further understand the nitrogen enrichment process.

In Fig.~\ref{fig:nitrogen_budget}, we track the mass evolution of the high-pressure ($>10^9\,\rm K\,cm^{-3}$) gas $m_{\rm highP}$, which is observed to be nitrogen-enriched~\citep[][and also found in our simulation]{PascaleDaiMcKee_2023ApJ...957...77P}. For three fiducial simulations at $Z=0.2\,Z_\odot$, $m_{\rm highP}$ can reach $\sim 10^5\,M_\odot$ at $t-t_0\sim 2\,\rm Myr$ (though this peak value slightly decreases for less compact GMCs), while it remains $m_{\rm highP}\sim 10^4\,M_\odot$ at $t-t_0\sim 4\,\rm Myr$ (i.e., the peak time of N/O anomaly, see Fig.~\ref{fig:fiducial_results}). However, the high-pressure gas is removed after $t-t_0\sim 4\,\rm Myr$.

With Eq.~\eqref{equ:no_fitting}, the initial mass fraction of nitrogen is $X_{\rm N,0} \approx 5.7\times 10^{-5}$, such that the primordial nitrogen mass in the high-pressure gas is $X_{\rm N,0}\,m_{\rm highP}$. This is a simple scale-down of $m_{\rm highP}$, and quickly drops to $\lesssim 1\,M_\odot$ after $t-t_0\sim 2\,\rm Myr$ (\emph{blue dashed line} of Fig.~\ref{fig:nitrogen_budget}). However, the total nitrogen mass in the high-pressure gas is $\sim 10\,M_\odot$. The nitrogen in the high-pressure gas therefore almost entirely originates from stellar winds after $\sim 3\,\rm Myr$ (\emph{blue shaded region}). Moreover, the fact that high-pressure gas is nitrogen-enriched due to stellar winds suggests that it must be in the proximity of stars to incorporate the wind material. 

Next, we consider nitrogen in the low-pressure ($p/k_{\rm B}<10^9\,\rm K\,cm^{-3}$) gas, which initially has a mass $X_{\rm N,0}\,M_{\rm cloud} \approx 1.71\times 10^3\,M_\odot$ (\emph{red asterisk}). This nitrogen mass first slightly decreases due to gas mass lost to star formation, but remains almost unchanged at $\sim 1.5\times 10^3\,M_\odot$ afterwards. Moreover, nitrogen expelled from stars accumulates with time and asymptotically reaches $\sim 200$--$300\,M_\odot$, significantly less than the primordial nitrogen mass. Therefore, unlike the nitrogen in the high-pressure gas which is largely from wind ejecta, the nitrogen in the low-pressure gas is dominated by the primordial component and cannot show dramatic N/O elevation.

At the peak time of N/O anomaly, the cumulative nitrogen expelled from stars amounts to $\sim 100\,M_\odot$, while the nitrogen in the high-pressure gas is only $\sim 10\,M_\odot$. This shows that most of the nitrogen from wind ejecta mixes into the low-pressure gas even at that time. This reinforces the realization that the high-pressure gas appears nitrogen-rich due to a high ratio between the captured wind mass and the natal gas mass, rather than it trapping most of the stellar nitrogen yield.

In summary, we find that the mass of the high-pressure gas is $\sim 10^4\,M_\odot$ at the culmination of nitrogen enrichment in the simulation, which is a tiny fraction of the total GMC mass ($3\times 10^7\,M_\odot$). Despite its insignificance in terms of mass, this gas is in very close proximity to massive stars at the center of the cluster, such that it may absorb a sizable or even dominant fraction of the stellar ionizing radiation, and hence dominate the observed nebular emission from the high-ionization zone. The two-component photo-ionization model in \cite{PascaleDaiMcKee_2023ApJ...957...77P} captures this. The high-pressure gas is a mixture of natal gas and, predominantly, stellar wind ejecta, such that most nitrogen there comes from nitrogen-rich winds of massive stars (also see Fig.~\ref{fig:toy_model_chemical}), particularly of the VMSs. While the high-pressure gas is indeed nitrogen-enriched, the majority of the gas in the GMC is ``low-pressure'' and lacks a dramatic N/O elevation, as also reflected in Fig.~\ref{fig:simulation_visualization}. Interestingly, \citet{JiUblerMaiolino_2024MNRAS.535..881J} found that the low-density gas of GS\_3073 is not enriched, similar to the prediction of the scenario here.

\subsection{Simulations with different GMC masses}
\label{sec:res:surface_density}

We run a group of control simulations with different GMC initial masses, while keeping the same GMC radii as in the fiducial simulations. The initial mean surface density of the GMCs will also change accordingly. 

Fig.~\ref{fig:results_mass_variation} shows the time evolution of star formation and the chemical abundances of the high-pressure gas for these control simulations (with the same convention as Fig.~\ref{fig:fiducial_results}). The final stellar mass reaches $\sim 5\times 10^7\,M_\odot$ for GMCs of $10^8\,M_\odot$ in contrast to $\sim 2\times 10^6\,M_\odot$ for GMCs of $10^7\,M_\odot$, showing that the star formation efficiency (SFE) increases (from $\sim 20\%$ to $\sim 50\%$) with the GMC mass, and more fundamentally, the GMC initial mean surface density as found by previous studies~\citep[e.g.,][and reference therein]{GrudicHopkinsFaucher-Giguere_2018MNRAS.475.3511G,ChevanceKrumholzMcLeod_2023ASPC..534....1C}.

In comparison to simulations with $M_{\rm cloud}=3\times 10^7\,M_\odot$, simulations with $M_{\rm cloud}= 10^8\,M_\odot$ show a significantly smaller UV half-light radius ($\sim 5\,\rm pc$). The half-mass radii of the central massive cluster, however, are quite similar to fiducial simulations. This may be the result of a violent relaxation process for the central cluster that eliminates part of the information about the initial condition. On the other hand, systems formed out of less massive GMCs ($M_{\rm cloud}= 10^7\,M_\odot$) still have small half-mass radii for the central cluster ($\sim 1 \,\rm pc$), but the UV half-light radius is much larger ($\sim 10-30\,\rm pc$), suggesting a larger stellar mass share of orbiting satellite sub-clusters and cluster outskirts.

The simulation \path{M1e8_R60_Z0.2} with the highest surface density for the natal GMC also has the smallest UV half-light radius ($\sim 2-3\,\rm pc$), which is close to the half-mass radius of the central cluster. We also visualize the stellar distribution of this simulation in the lower panels of Fig.~\ref{fig:star_cluster_scattering}, where almost all stars belong to the central massive cluster (though there are indeed many smaller satellite clusters). All other simulations typically resemble the upper panels of Fig.~\ref{fig:star_cluster_scattering}, where sub-clusters are gravitationally unbound from each other and scatter throughout the entire simulation domain (so the UV half-light radius of the system is much larger than the half-mass radii of sub-clusters). The simulation \path{M1e8_R60_Z0.2} is an example of hierarchical mergers where sub-clusters are gravitationally bound to the central cluster that they will eventually merge into~\citep{ShiKremerHopkins_2024ApJ...969L..31S}. 

We further check the chemical evolution of the high-pressure ($>10^9\,\rm K\,cm^{-3}$) gas in these two sets of simulations. Simulations with $M_{\rm cloud}= 10^8\,M_\odot$ have similar N/O and C/O signals to those of the fiducial simulations, but enrichment occurs earlier, in only $\sim 2-3\,\rm Myr$. However, the high-pressure gas in simulations with $M_{\rm cloud}= 10^7\,M_\odot$ can only rise to $\log(\rm N/O) \sim -1$ throughout the evolution, well below the N/O value found in \citet{PascaleDaiMcKee_2023ApJ...957...77P}. As a result, these less massive and less compact GMCs are unlikely to be the progenitor of the LyC cluster. In fact, less compact GMCs may not even confine gas with $p/k_{\rm B}>10^9\,{\rm K}\,{\rm cm}^{-3}$ in its star-forming complex. A simple, order-of-magnitude argument is that the self-gravity, $g\sim G M_{\rm cloud}/R_{\rm cloud}^2 \sim \pi G \Sigma$, should be greater than the buoyancy, $\nabla p /\rho \sim (p/R_{\rm cloud}) / (3M_{\rm cloud}/4\pi R_{\rm cloud}^3) \sim p /(3\Sigma /4)$. Insert $p= k_{\rm B} \cdot 10^9\,{\rm K}\,{\rm cm}^{-3}$, we find $\Sigma \gtrsim (4p/3\pi G)^{1/2} \gtrsim 1\,{\rm g}\,{\rm cm}^{-2}$, roughly corresponds to the clouds we are simulating.

\begin{figure*}
    \centering
    \includegraphics[width=\linewidth]{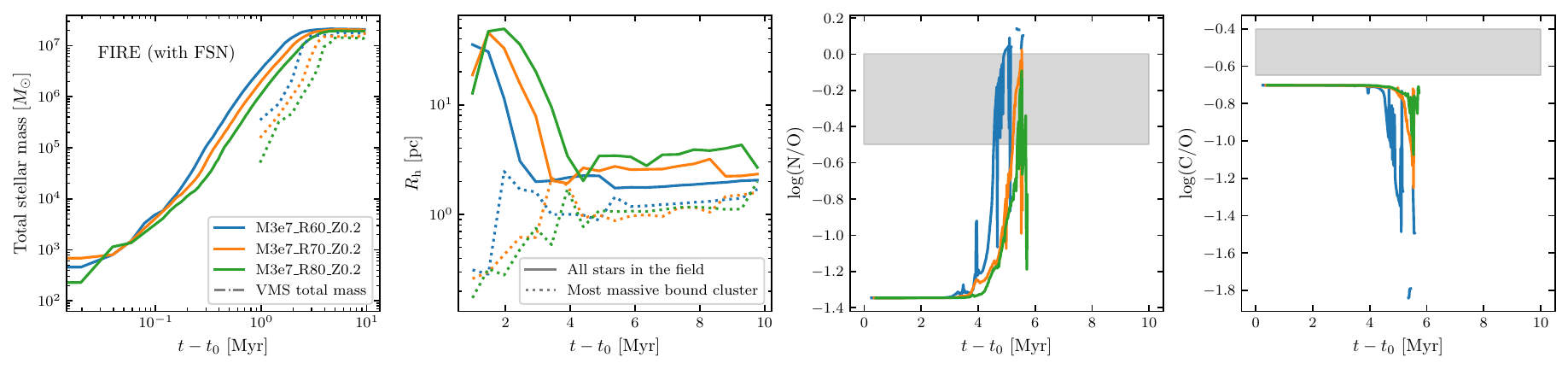}
    \includegraphics[width=\linewidth]{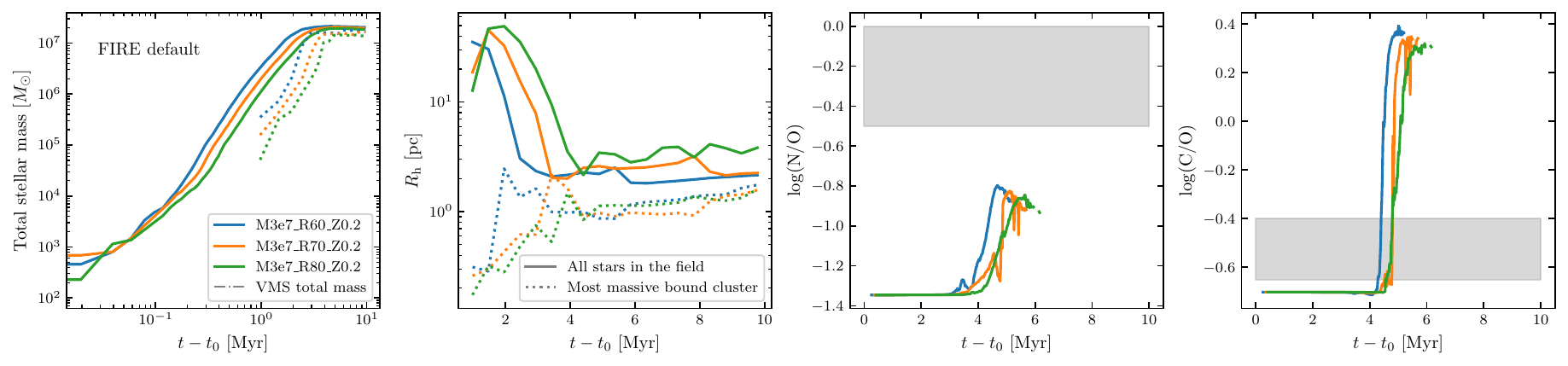}
    \caption{Simulations without VMSs (i.e., FIRE-only). Conventions are the same as Fig.~\ref{fig:fiducial_results}. \emph{Top row} -  FIRE simulations but consider the ``failed'' SNe (corresponding to ZAMS mass of $\sim 25-70\,M_\odot$, see \S~\ref{sec:sim:fire}). \emph{Bottom row} - simulations with default FIRE setups. 
    }
    \label{fig:fire_sims}
\end{figure*}

\subsection{Simulations without VMSs}
\label{sec:res:fire}

To consolidate the argument that VMSs are \emph{required} for strong N/O elevation in newborn super star clusters, we run FIRE-only simulations \emph{without} VMSs and summarize the results in Fig.~\ref{fig:fire_sims}. 

In these simulations, chemical feedback is completely determined by the sub-grid models of FIRE-3~\citep{HopkinsWetzelWheeler_2023MNRAS.519.3154H}. We run two versions of FIRE-only simulations. First, we consider ``failed SNe'' such that there is no SN feedback before $t_{\rm CC}\sim 8\,\rm Myr$. This is also the modification that we made for our FIRE+VMS model described earlier (\S~\ref{sec:sim:fire}), but the difference is that no VMSs are modeled and the IMF extends only to $100\,M_\odot$ (so PPISNe at $\sim 3\,\rm Myr$ are also missing). This leads to weaker stellar feedback. This is reflected in the star formation history, where the stellar mass accumulates to $\sim 2\times 10^7\,M_\odot$, roughly twice as much compared with the FIRE+VMS runs (\S~\ref{sec:res:fiducial}). 

This is consistent with the theory~\citep{MurrayQuataertThompson_2010ApJ...709..191M,FallKrumholzMatzner_2010ApJ...710L.142F,MatznerJumper_2015ApJ...815...68M} and simulations~\citep{GrudicHopkinsFaucher-Giguere_2018MNRAS.475.3511G,ChevanceKrumholzMcLeod_2023ASPC..534....1C} that the SFE is determined by the competition between self-gravity, $F_{\rm g} \sim G\, M_{\rm cloud} \,M_{\rm gas}/R_{\rm cloud}^2 \sim \pi\,G \,M_{\rm gas}\,\Sigma_{\rm cl}$, and the feedback force, characterized by $F_{\rm fb} \sim (\dot p / m) M_\star$, where $\dot p/m$ is the momentum ejection rate per unit mass from stars. By removing the VMS populations, $\dot p /m$ for the stellar population decreases since VMSs have a higher L/M ratio, which leads to the dominance of self-gravity and a higher SFE in the new balance. The dominance of self-gravity also explains the UV half-light radii of the FIRE-only runs, which are $\sim 2-3\,\rm pc$ after $t-t_0 \sim 4\,\rm Myr$, significantly smaller than in simulations with VMSs (Fig.~\ref{fig:fiducial_results}).

Another version of the FIRE-only runs keeps all the original setups of FIRE-3. We find no significant difference in terms of star formation when compared to the FIRE-only runs with ``failed SNe.'' SN feedback takes effect when the age of the stellar population (i.e., $t-t_0$) reaches $3.7\,\rm Myr$~\citep{HopkinsWetzelWheeler_2023MNRAS.519.3154H}, too late to impact star formation.

The differences between the two simulation versions are more significant in chemical feedback. With FSN implemented, chemical feedback before $t_{\rm cc}\sim 8\,\rm Myr$ in the FIRE-only runs is dominated by stellar winds from stars with $<100\,M_\odot$. We indeed see nitrogen elevation to $\log(\rm N/O) \gtrsim -0.5$ at $t-t_0\sim 5\,\rm Myr$, compatible with our toy model compiled with PARSEC data (Fig.~\ref{fig:toy_model_chemical}), but the timing of this nitrogen enrichment event $\sim 5\,\rm Myr$ is later than the observed $2-4\,\rm Myr$ range~\citep{PascaleDaiMcKee_2023ApJ...957...77P}. Moreover, the C/O ratio decreases with time, down to $\sim -1.4$ at $t-t_0\sim 5\,\rm Myr$, which becomes inconsistent with observations. Like in the fiducial simulations (Fig.~\ref{fig:fiducial_results}), high-pressure gas is removed after nitrogen enrichment, but this takes place later at $t-t_0\sim 6\,\rm Myr$. Still, this is before the onset of CCSNe in this simulation, ruling out the possibility that the high-pressure gas is removed by SN feedback.

The default FIRE-3 simulations do not reproduce the chemical feedback either. Although there is nitrogen enrichment at $t-t_0\sim 5\,\rm Myr$, it does not reach the observed range. Given the difference between the previous simulations with FSN, the nitrogen enrichment should be ``interrupted'' by oxygen yield from SN explosions. The C/O ratio here, however, surges to $\log({\rm C/O})>0$, which is also consistent with observations. This specific result is related to the SN yield model of FIRE-3~\citep{HopkinsWetzelWheeler_2023MNRAS.519.3154H}, where a high abundance of carbon is produced in early SN explosions.

Running these simulations without VMSs, we find that the expected nitrogen enrichment in the time window $2$--$4\,\rm Myr$ cannot be well reproduced. This demonstrates that VMSs play a key role in this process in our simulation. This is also reflected in similar simulations presented by \citet{FukushimaYajima_2024PASJ...76.1122F} where the IMF extends to $120\,M_\odot$ and $\log(\rm N/O)$ is enhanced by at most $\sim 0.75\,\rm dex$, potentially due to the insufficient nitrogen mass budget from the canonical IMF.

\begin{figure*}
    \centering
    \includegraphics[width=\linewidth]{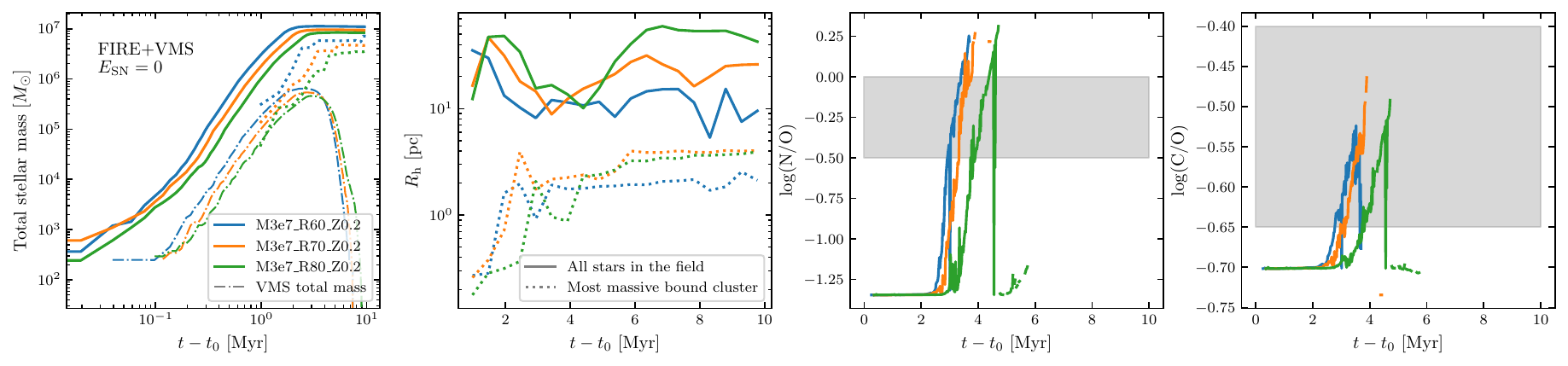}
    \includegraphics[width=\linewidth]{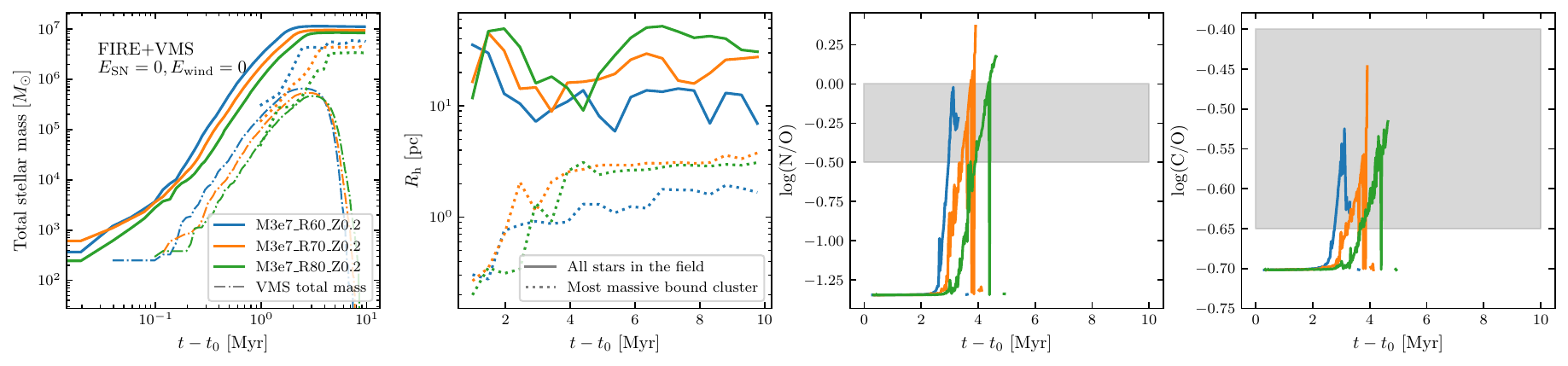}
    \includegraphics[width=\linewidth]{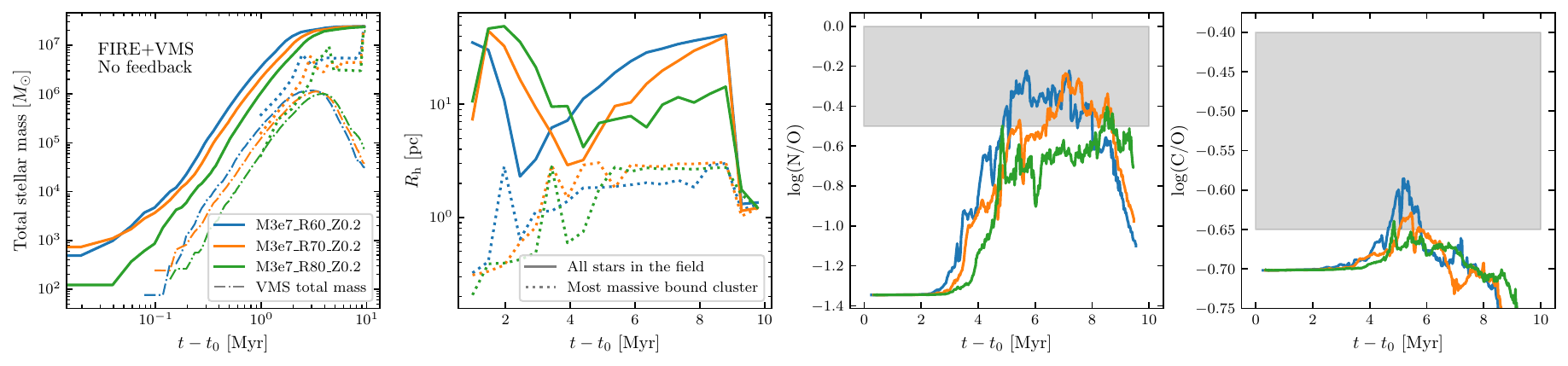}
    \caption{Simulations to test the feedback from VMSs. Here, the simulations keep the fiducial setups but remove the energy of VMS-origin SNe (\emph{first row}), or additionally the kinetic energy of VMS winds (by reducing the wind velocity; \emph{second row}), or no feedback (by also removing the radiative feedback; \emph{third row}). The feedback from FIRE stellar populations is unaltered for the first two tests, but is also disabled in the third one. Conventions are the same as Fig.~\ref{fig:fiducial_results}.
    }
    \label{fig:vms_feedback_tests}
\end{figure*}

\subsection{VMS feedback tests}
\label{sec:res:feedback_tests}

Figure~\ref{fig:high_no} shows that nitrogen-enriched gas in the simulation rapidly sinks into the central cluster before being expelled. To identify the feedback mechanism responsible (radiation, winds, or SNe), we rerun the simulations with specific feedback channels disabled. This is achieved by reducing the associated energy by a factor of $10^{-10}$ ($10^{-5}$ for wind momentum) while keeping the mass outflow and chemical feedback unchanged. We perform three tests: 
(1)~VMS SNe with kinetic feedback suppressed ($E_{\rm SN}\approx0$); 
(2)~both VMS SNe and wind kinetic feedback disabled ($E_{\rm SN}\approx E_{\rm wind}\approx0$); and 
(3)~all stellar radiative feedback turned off (for both FIRE SSPs and VMSs), leaving only the weak winds from FIRE populations until $\sim8\,\rm Myr$. 
Results for these runs are shown in Fig.~\ref{fig:vms_feedback_tests}, following the format of Fig.~\ref{fig:fiducial_results}.

For simulations with $E_{\rm SN} \approx 0$ or with $E_{\rm SN} \approx E_{\rm wind} \approx 0$ (i.e., first two rows in Fig.~\ref{fig:vms_feedback_tests}), we find no qualitative difference compared to the fiducial simulations (Fig.~\ref{fig:fiducial_results}). The star formation histories of these simulations are very similar. While chemical enrichment of high-pressure gas (which still exists even though the kinetic outputs from SNe and winds are turned off) shows some difference, e.g., nitrogen enrichment in the \path{M3e7_R70_Z0.2} simulation is delayed when kinetic outputs of SNe and winds are both disabled, the overall degree of N/O elevation and the timing to reach the observed range are still largely the same. 

Therefore, the only remaining strong feedback mechanism has to do with radiation. In the third test, we disable the radiative feedback from all stars (the VMS population alone elevates the overall L/M ratio, but only by a factor of $\sim 1.5$, see \S~\ref{sec:sim}). With the third row in Fig.~\ref{fig:vms_feedback_tests}, we see that radiative feedback indeed makes a big difference. The SFE is much higher as the cumulative stellar mass reaches $\sim 2.5\times 10^7\,M_\odot$, which amounts to $\sim 80\%$ of the GMC mass converted into stars.

More importantly, by showing the N/O and C/O abundance ratios of the high-pressure gas (right two panels), we find that $\log (\rm N/O)$ can still exceed $-0.5$ but happens at a delayed time at $\gtrsim 5\,\rm Myr$, which is later than the inferred age of the Sunburst LyC cluster. There is also a trend that less compact GMCs tend to have more delayed nitrogen enrichment and lower $\rm N/O$ ratios. Another important difference from the previous two feedback tests is that the high-pressure gas persists till the termination of simulation. By comparison, that gas is removed after $\sim 4$--$5\,\rm Myr$ in simulations with radiative feedback. Stellar radiation has a key impact on the dynamic evolution of the nitrogen-enriched gas. During gravitational collapse of the gas early on, radiation heats the gas and injects momentum into it. This pressurizes the gas but also renders the high-pressure gas dynamics less stable. 

We conclude that radiative feedback plays an important dual role in shaping the nitrogen enrichment process -- without radiation, the natal gas cannot be efficiently removed from the star-forming region, so enrichment is diluted until significantly later (akin to the ``closed box'' model in \S~\ref{sec:toy_model}). Stellar winds supply chemical elements, but are not the primary source of momentum feedback. Finally, SN explosions do not affect the enrichment process either, primarily because chemical enrichment occurs too rapidly at $2$--$4\,\rm Myr$, and many VMSs terminate their evolution within this period as failed SNe. However, kinetic feedback from SNe may become important once the age of the stellar population exceeds $t_{\rm CC}\sim 8\,\rm Myr$.

\section{Discussion}
\label{sec:discussion}

\subsection{Wolf-Rayet stars and stellar metallicity}

\citet{TapiaBekkiGroves_2024MNRAS.534.2086T} modeled the origin of nitrogen anomaly of the Sunburst LyC cluster with OB stars and found that a young cluster can be nitrogen-enriched at an age $4.8$--$5.4\,\rm Myr$, when stellar winds mix with about $\sim 1\%$ leftover natal gas. However, we note that this timing of nitrogen enrichment appears somewhat later than the measured cluster age~\citep{Rivera-ThorsenChisholmWelch_2024A&A...690A.269R}, as too much leftover natal gas would greatly dilute the enrichment signal. Similarly, our FIRE-only simulations (where the IMF does not extend beyond $100\, M_\odot$) also suggest that the N/O ratio reaches the observed range only after $\sim 5\,\rm Myr$ (Fig.~\ref{fig:fire_sims}). As a result, a normal IMF without VMSs may not be able to explain the fast ($<4\,\rm Myr$) enrichment of nitrogen.

Moreover, based on the C/O abundance ratio, \citet{TapiaBekkiGroves_2024MNRAS.534.2086T} excluded the presence of WR stars in the system to avoid over-production of carbon, while this seems incompatible with the detection of WR spectral features~\citep{Rivera-ThorsenChisholmWelch_2024A&A...690A.269R}. From PARSEC data~\citep[][see \S~\ref{sec:toy_model}]{CostaShepherdBressan_2025A&A...694A.193C}, we found that the yields of both carbon and nitrogen have non-monotonic metallicity dependence. In Fig.~\ref{fig:toy_model_chemical}, we also present the chemical composition of the stellar winds, where the C/O ratio can exceed $\log(\rm C/O)=0$ after $\sim 3\,\rm Myr$ for most metallicities in the plot except $Z=0.002$ and $Z=0.004$ (note that the inferred metallicity of the LyC cluster is in this range). As a result, the observed level of C enrichment by WR stars is reproduced by our simulations, as WR stars can exist in the cluster without depositing too much carbon due to the metallicity dependence of chemical yields. Similarly, \citet{MollaTerlevich_2012MNRAS.425.1696M} calculated the cumulative element yields of stellar populations for different metallicity values. They found that the carbon abundance ($12+\log(\rm C/H)$, see their Fig.~9) in stellar winds is strongly enriched for $Z\ge 0.08$, but is actually slightly decreased for $Z\le 0.004$. These results suggest that WR winds do not overproduce carbon at intermediate metallicity. 

\citet{MollaTerlevich_2012MNRAS.425.1696M} also found that N/H is enhanced by $\sim 1$ dex due to stellar winds for all metallicities (while the enhancement in C/H only appears for $Z>0.2\,Z_\odot$). The oxygen abundance, however, can grow by $\sim 1$ dex for $Z=0.008$ and $Z=0.02$, but only slightly declines with time for $Z\le 0.004$. This implies that a high N/O due to stellar winds is only possible for $Z\le 0.004$. We note that this behavior is in agreement with our compilation based on \citet{CostaShepherdBressan_2025A&A...694A.193C} in \S~\ref{sec:toy_model}.

Interestingly, from a comprehensive compilation of the nitrogen-enriched star clusters and high-redshift galaxies~\citep[][see their Table 2]{JiBelokurovMaiolino_2025MNRAS.tmp.1982J}, the metallicities of these objects (including the Sunburst LyC cluster) are typically around $12+\log(\rm O/H)\sim 7.5-8$~\citep[also see $z<0.5$ N-emitters from DESI DR1,][]{BhattacharyaKobayashi_2025arXiv250811998B}. textcolor{black}{More recent data of N-emitters from \citet{MorelSchaererMarques-Chaves_2025arXiv251120484M} revealed a wider range of $12+\log(\rm O/H)\sim 7-8.5$, yet still sub-solar. Moreover, the C/O abundance of these N-emitters (including the Sunburst Arc, see Fig.~\ref{fig:no_co_fittings}) aligns with the local relation.} This could be a reflection of the intrinsic metallicity-dependent stellar evolution that affects the element yields of massive star winds~\citep{MollaTerlevich_2012MNRAS.425.1696M}.

We suggest that the trend arises from metallicity-dependent yields during the WR phase of massive stars and VMSs~\citep{Maeder_1992A&A...264..105M}. During this phase, helium burning in the core (via the triple-alpha reaction) produces carbon and oxygen, while hydrogen burning in the shell (through the CNO cycle) enriches nitrogen. At high metallicity, strong winds strip first the envelope and then the core, driving the spectral evolution from WN to WC/WO stars and ejecting substantial nitrogen, carbon, and oxygen~\citep{Maeder_1992A&A...264..105M,HigginsVinkHirschi_2024MNRAS.533.1095H}. At low metallicity, weaker winds~\citep{TramperSanadeKoter_2016ApJ...833..133T} remove mainly the nitrogen-rich envelope but rarely the core~\citep{Aguilera-DenaLangerAntoniadis_2022A&A...661A..60A}, making WC/WO stars rarer~\citep{EldridgeVink_2006A&A...452..295E,Crowther_2007ARA&A..45..177C} and yielding winds with high N/O and little carbon.

In summary, WR stars and VMSs likely formed in the Sunburst LyC cluster. The SMC-like metallicities observed in nitrogen-enriched galaxies and clusters~\citep[][]{JiBelokurovMaiolino_2025MNRAS.tmp.1982J} are consistent with WR evolutionary models~\citep{Aguilera-DenaLangerAntoniadis_2022A&A...661A..60A}.

\subsection{High-redshift nitrogen-enriched galaxies}

Recent advances in observations, especially with JWST, have uncovered that a small fraction (a few percent) of high-redshift ($z\sim 5$--$11$) galaxies exhibit strong nebular emission lines indicating an elevated gas-phase nitrogen abundance~\citep[e.g.,][]{CameronKatzRey_2023MNRAS.523.3516C,BunkerSaxenaCameron_2023A&A...677A..88B,IsobeOuchiTominaga_2023ApJ...959..100I,Navarro-CarreraCaputiIani_2025ApJ...993..194N,SchaererMarques-ChavesXiao_2024A&A...687L..11S,Marques-ChavesSchaererKuruvanthodi_2024A&A...681A..30M,JiUblerMaiolino_2024MNRAS.535..881J,ToppingStarkSenchyna_2025ApJ...980..225T}. These ``N-emitters'' feature high N/O ratios, with some even reaching $\log(\rm N/O)\gtrsim 0.5$~\citep[e.g.,][]{IsobeOuchiTominaga_2023ApJ...959..100I,JiUblerMaiolino_2024MNRAS.535..881J}. Moreover, N-emitters can be associated with starbursts~\citep{Alvarez-MarquezCrespoGomezColina_2025A&A...695A.250A}.

We speculate that the physical processes localized to newborn super star clusters as discovered in this work could be responsible for the spectral behavior of these high-redshift N-emitters. VMSs are theoretically expected to exist in the early Universe~\citep{SchaererGuibertMarques-Chaves_2025A&A...693A.271S}, and at least an excess of $>50\,M_\odot$ massive stellar population has been observed~\citep{CullenCarnallScholte_2025MNRAS.540.2176C}. Once starbursts happen in galaxies, the newly formed young stars may enrich nearby gas through stellar winds. The enriched gas may not be representative of the mean ISM properties of the whole galaxy, but instead only a small amount of high-pressure, dense photo-ionized gas in the proximity of highly massive, compact star clusters. This can explain why the nitrogen-enriched nebulae are compact in size~\citep[e.g.,][]{ToppingStarkSenchyna_2024MNRAS.529.3301T,ToppingStarkSenchyna_2025ApJ...980..225T}. We also note that several observations~\citep{IsobeOuchiTominaga_2023ApJ...959..100I,SenchynaPlatStark_2024ApJ...966...92S,JiBelokurovMaiolino_2025MNRAS.tmp.1982J} found chemical similarities between N-emitters and nitrogen-rich globular cluster stars,
suggesting possibly an underlying connection between these two contexts. 

\cite{McClymontTacchellaSmith_2025arXiv250708787M} attribute ISM nitrogen enhancement in galaxies to recent starbursts and AGB stars. 
Similarly, chemical evolution models by \citet{RizzutiMatteucciMolaro_2025A&A...697A..96R} found that differential winds can lead to the enrichment, while oxygen and $\alpha$ elements are produced by CCSNe. These different scenarios, as well as the VMS scenario presented here, may need further detailed studies to confirm in the context of high-redshift N-emitters. However, both AGB stars and CCSNe are typically much older than $4\,\rm Myr$, so we argue that they are unlikely to explain the nitrogen enrichment of the Sunburst Arc LyC cluster. The winds are also unlikely to be as dense as the observed density of $\gtrsim 10^5\,\rm cm^{-3}$ for some high-redshift N-emitters~\citep{PascaleDaiMcKee_2023ApJ...957...77P,JiUblerMaiolino_2024MNRAS.535..881J,YanagisawaOuchiWatanabe_2024ApJ...974..266Y}.

Finally, from our simulation, star clusters here can have high central densities ($\sim 10^5\, M_\odot\,{\rm pc}^{-3}$), so another potential source in the nitrogen mass budget may be supermassive stars~\citep{CharbonnelSchaererPrantzos_2023A&A...673L...7C,NandalWhalenLatif_2025ApJ...994L..11N,GielesPadoanCharbonnel_2025MNRAS.544..483G} as the product of runaway stellar mergers and the transition to intermediate-mass black holes~\citep[IMBHs; e.g.,][]{PortegiesZwartMcMillan_2002ApJ...576..899P,GurkanFreitagRasio_2004ApJ...604..632G,KremerSperaBecker_2020ApJ...903...45K,ShiGrudicHopkins_2021MNRAS.505.2753S,FujiiWangTanikawa_2024Sci...384.1488F}. However, the relative importance of this channel may require a closer look given that VMSs alone can already explain the N/O anomaly. Particularly for the Sunburst Arc, the VMSs and their mass segregation are detected~\citep{MestricVanzellaUpadhyaya_2023A&A...673A..50M}, while a supermassive star should feature a central bright point source, so we argue that supermassive stars may not be the explanation for this object.

\begin{figure}
    \centering
    \includegraphics[width=\linewidth]{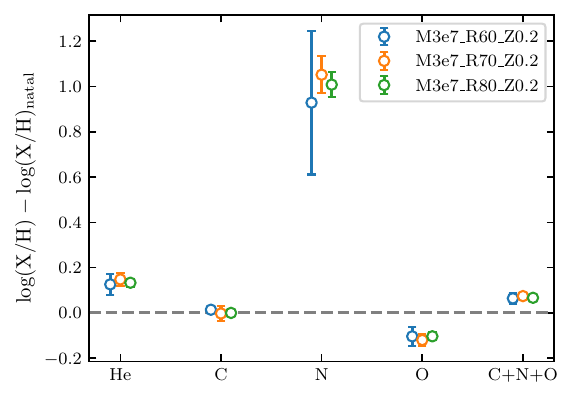}
    \caption{The abundance ratio between element ``X'' (as shown on the horizontal axis) and hydrogen for the high-pressure ($\gtrsim 10^9\,\rm K\, cm^3$), nitrogen-enriched gas (selected at the first snapshot that reaches $\log(\rm N/O)\ge -0.25$) taken from our fiducial simulations.
    }
    \label{fig:x_over_h}
\end{figure}

\subsection{Nitrogen-enriched AGNs}

Observations also found high-redshift, nitrogen-enriched broad-line AGNs~\citep{IsobeMaiolinoD'Eugenio_2025MNRAS.541L..71I}, which can also be related to this scenario, that the broad-line region (BLR) is polluted by ``field stars'' in the nuclear star cluster after a starburst. However, we caution that the source of this enrichment can also be related to star formation inside AGN disks, which undergoes a completely different evolution track compared with field stars~\citep{CantielloJermynLin_2021ApJ...910...94C,JermynDittmannMcKernan_2022ApJ...929..133J,Ali-DibLin_2023MNRAS.526.5824A}. For example, \citet{HuangLinShields_2023MNRAS.525.5702H} found that apart from the nitrogen enrichment in the BLR, there is also significant C+N+O enrichment. This differs from the globular cluster situation where C+N+O is roughly uniform among multiple stellar populations~\citep[which is the case for narrow-line region;][]{HuangLinShields_2023MNRAS.525.5702H}, and hence implies a different chemical feedback process~\citep{FryerHuangAli-Dib_2025MNRAS.537.1556F}.

Fig.~\ref{fig:x_over_h} shows the abundance of different elements of the high-pressure, nitrogen-enriched gas from our fiducial star cluster simulations (\S~\ref{sec:res:fiducial}). The abundances of C and O exhibit unremarkable deviations from the natal values (at most $\sim 0.2\,$ dex). Apart from the significant N enrichment, it is interesting to see a significant elevation in He abundance ($0.1$--$0.2\,$dex). However, the sum C+N+O is very similar to the natal value, which is unlike the case of BLR enrichment in \citet{HuangLinShields_2023MNRAS.525.5702H}. VMSs may or may not play a role in the nitrogen enrichment of AGNs.

\section{Conclusion}
\label{sec:conclusion}

By semi-analytic calculations (\S~\ref{sec:toy_model}) and running magnetohydrodynamical simulations with radiative feedback (\S~\ref{sec:sim}, \ref{sec:res}), we have presented evidence that nitrogen enrichment of the high-pressure gas in the Sunburst LyC cluster is the combined result of both the formation of \emph{massive compact star cluster} and chemical feedback from \emph{very massive stars}. The enrichment source can be a hybrid of both \citet[][VMSs only]{Vink_2023A&A...679L...9V} and \citet[][WR stars with non-VMS origin]{KobayashiFerrara_2024ApJ...962L...6K}. But a key difference is that the effect we study operates on the star cluster scales rather than galaxy scales, given the observed stellar wind signatures of VMSs and/or WR stars~\citep{MestricVanzellaUpadhyaya_2023A&A...673A..50M,Rivera-ThorsenChisholmWelch_2024A&A...690A.269R}.
We summarize below the key physical processes of cluster formation and feedback:

\begin{enumerate}
    \item \emph{Super star cluster formation from dense GMC collapse}. For an initial GMC with a mean surface density as high as $\sim 10^3$--$10^4\,M_\odot\,{\rm pc}^{-2}$, a star cluster may form at a very high star formation efficiency of $\sim 30\%$, leading to a super star cluster (consisting of a central cluster and many satellites) of $\sim 10^7\,M_\odot$ that matches the inferred stellar mass of the Sunburst LyC cluster~\citep{VanzellaCastellanoBergamini_2022A&A...659A...2V,PascaleDaiMcKee_2023ApJ...957...77P,Rivera-ThorsenChisholmWelch_2024A&A...690A.269R}. If the initial GMC is more compact, hierarchical mergers build up an even more mass-dominating central cluster~\citep{ShiKremerHopkins_2024ApJ...969L..31S}. The compact star cluster can confine the gas in its cluster.
    
    \item \emph{Gas sinking under the pull of gravity}. Within the first global free-fall time of the GMC, cold natal gas is pulled in by gravity toward the cluster center. The density of this gas dramatically increases ($\gtrsim 10^5\,\rm cm^{-3}$) through this nearly isothermal gravitational collapse. 
    We identify this as a new dynamical reason to account for the observation that nitrogen-enriched gas is denser (and more pressurized) than typical warm ionized gas in the ISM~\citep[e.g.,][]{PascaleDaiMcKee_2023ApJ...957...77P,JiUblerMaiolino_2024MNRAS.535..881J,YanagisawaOuchiWatanabe_2024ApJ...974..266Y}.
    
    \item \emph{Chemical yields of VMS winds}. Stellar winds from massive stars, especially from VMSs, are the source of pollution that enriches the dense gas pulled to the cluster center. Particularly for the Sunburst LyC cluster, the VMS population at $Z=0.2\,Z_\odot$ enrich the natal gas that has fallen to the center, such that $\log(\rm N/O)\gtrsim -0.5$ while carbon is not over-abundant. This matches what is measured from the observed high-pressure nebula~\citep{PascaleDaiMcKee_2023ApJ...957...77P}. We also predict that the dense gas exhibits $0.1$--$0.2\,$dex of helium abundance elevation.
    
    \item \emph{Stellar irradiation.} Simultaneously with chemical enrichment, the dense gas becomes photo-heated and ionized, and is subject to strong radiation pressure, while kinetic feedback from winds and SNe is secondary (\S~\ref{sec:res:feedback_tests}). Radiation pressure therefore is an important contributor to the high pressure of the nitrogen-enriched gas, while most of the remaining natal gas at larger radii has a low pressure and is only slightly enriched~\citep{PascaleDaiMcKee_2023ApJ...957...77P}~\citep[i.e., stratification of nebular physical and chemical properties;][]{PascaleDaiMcKee_2023ApJ...957...77P, JiUblerMaiolino_2024MNRAS.535..881J, ToppingSandersShapley_2025MNRAS.541.1707T}. Soon after enrichment of the innermost dense gas, which amounts to $\sim 10^4\,M_\odot$, strong irradiation efficiently ejects it out of the cluster potential, at around $\sim 4\,\rm Myr$. 
\end{enumerate}

\emph{Massive compact star cluster formation} requires natal GMCs with a high surface density. Our fiducial simulations suggest that GMCs weighing $3\times 10^7\,M_\odot$ and initially of a size $\sim 70\,\rm pc$ (mean surface density $\Sigma\sim 2\times 10^3\,M_\odot\,{\rm pc}^{-2}$) can collapse to give the desired cluster mass and radius as observed for the Sunburst LyC cluster, while simulations with lower initial surface density values do not due to a lower star formation efficiency~\citep{GrudicHopkinsFaucher-Giguere_2018MNRAS.475.3511G,ChevanceKrumholzMcLeod_2023ASPC..534....1C}. This requirement of high surface density agrees with other simulation results that focus on the escape of LyC photons from massive compact star clusters~\citep{MenonBurkhartSomerville_2025ApJ...987...12M}. Moreover, what forms out of the natal GMC is not a single star cluster, but instead a central massive cluster surrounded by many satellite sub-clusters (Fig.~\ref{fig:star_cluster_scattering}).

\emph{Very Massive Stars} are required to explain the rapid, heavy enrichment of nitrogen in the Sunburst LyC cluster at $\sim 2$--$4\,\rm Myr$. Though OB stars might lead to similar nitrogen enrichment, the time required to reach the observed high N/O value would be too long compared to observations, and this scenario would also require the absence of WR stars~\citep{TapiaBekkiGroves_2024MNRAS.534.2086T}. This is in tension with spectroscopy~\citep{MestricVanzellaUpadhyaya_2023A&A...673A..50M,Rivera-ThorsenChisholmWelch_2024A&A...690A.269R}.

\emph{Wolf-Rayet stars}, specifically WN-type stars, as a late and helium-burning stage of massive stars, and VMSs, should either be present or have formed before in the LyC cluster~\citep{Rivera-ThorsenChisholmWelch_2024A&A...690A.269R}. At the inferred metallicity values of nitrogen-enriched galaxies and super star clusters~\citep[$12+\log (\rm O/H)\sim 7.5-8$, or $Z\lesssim 0.2\,Z_\odot$, see][]{JiBelokurovMaiolino_2025MNRAS.tmp.1982J}, stellar winds efficiently shed the N-rich envelope of WN stars while carbon and oxygen are trapped in the stellar core~\citep{Aguilera-DenaLangerAntoniadis_2022A&A...661A..60A}, leading to a high N/O but slightly elevated C/O in the wind ejecta. Both our yield model (\S~\ref{sec:toy_model}) and \citet{MollaTerlevich_2012MNRAS.425.1696M} find this metallicity dependence in terms of N/O and C/O ratios.

We suggest that in high-redshift nitrogen-enriched galaxies like GN-z11 and GS\_3073, N/O elevation may represent a temporary phase during the bursty formation of dense stellar systems. Compared to the scenario of differential winds~\citep{RizzutiMatteucciMolaro_2025A&A...697A..96R,McClymontTacchellaSmith_2025arXiv250708787M}, this is associated with gas trapped in the proximity of newborn super star clusters~\citep[galaxies with N/O anomaly do have high surface density $\Sigma_\star \gtrsim 10^{2.5}\,M_\odot\,{\rm pc}^{-2}$, see][]{JiBelokurovMaiolino_2025MNRAS.tmp.1982J}. The VMS population provides the extra nitrogen, or more generally speaking, WR stars at sub-solar metallicity (Fig.~\ref{fig:toy_model_chemical}). This enrichment, occurring in dense gas that is pressurized and photoionized by young stars, can explain the observed nitrogen nebular emission lines. By contrast, the majority of the gas is low-density, low-pressure gas, which is not enriched \citep{JiUblerMaiolino_2024MNRAS.535..881J}.

One prediction of this scenario is a significant, $0.1$--$0.2\,$dex elevation of the He/H ratio in the high-pressure gas that would accompany the N/O anomaly (see Fig.~\ref{fig:simulation_visualization}), which can be measurable and explain the positive He/H-N/O correlation~\citep{YanagisawaOuchiWatanabe_2024ApJ...974..266Y}. This enrichment is also due to winds from massive stars including the VMSs, when the envelope of these stars becomes helium-elevated in their WR stages. Similar to the case of the N/O abundance ratio, the majority of the gas, being low-pressure, is more dominated by the natal gas components and is hardly helium-enhanced.

The process studied here may be related to the problem of multiple stellar populations in globular clusters, i.e., significant star-to-star variations in abundances of light elements like He, C, N, O, Na, and Al \citep{BastianLardo_2018ARA&A..56...83B,MiloneMarino_2022Univ....8..359M}. We note that stellar winds from VMSs, as described above, can mix effectively with the pristine gas in the deep potential well of the super star cluster, and the mixture can give birth to a second generation of stars, leaving a unique chemical imprint.

There are several caveats to this study.  The PARSEC tracks used here assume non-rotating stars~\citep{CostaShepherdBressan_2025A&A...694A.193C}, while stellar rotation introduces uncertainties in the yield model. textcolor{black}{The detailed physics of VMS modeling, like the mass loss recipes, can be complicated and metallicity-dependent \citep{SanderVink_2020MNRAS.499..873S,SabhahitVinkSander_2023MNRAS.524.1529S}. Another uncertainty is the VMS wind velocity, since observations found a fast terminal velocity of $\sim 3000\,\rm km/s$ for VMSs in the R136 clusters \citep{BrandsdeKoterBestenlehner_2022A&A...663A..36B}, unlike the slow-wind model \citep{Vink_2018A&A...615A.119V} used here. All these factors may change the result of the present study.} The simulations employ FIRE-3 physics with a presumed IMF for computational efficiency, whereas fully self-consistent frameworks such as STARFORGE~\citep{GrudicGuszejnovHopkins_2021MNRAS.506.2199G} would better capture the formation of single stars. Observations suggest that the number of VMSs inferred spectroscopically is lower than predicted by a standard IMF~\citep{MestricVanzellaUpadhyaya_2023A&A...673A..50M}, highlighting the need for a more realistic IMF in these environments (or reflecting the rapid evolution of VMSs). Our simulations also lack the resolution to model turbulent mixing and cooling between hot shocked winds and cool photoionized gas~\citep[e.g.,][]{JiOhMasterson_2019MNRAS.487..737J}, which may affect the dynamics of enriched dense gas near cluster centers. Additional physics, such as runaway stellar mergers, are likewise omitted. We leave these to future work.

\begin{acknowledgments}
We thank Kendall Shepherd and Emil Rivera-Thorsen for useful comments. Y.S., N.M., and C.S.Y. acknowledge the support of the Natural Sciences and Engineering Research Council of
Canada (NSERC) under the funding reference number 568580. L.D. acknowledges research grant support from the Alfred P. Sloan Foundation (Award Number FG-2021-16495) and from the Frank and Karen Dabby STEM Fund in the Society of Hellman Fellows.
Part of the computation is performed at the University of Toronto cluster ``Niagara,''  supported by SciNet (scinethpc.ca) and the Digital Research Alliance of Canada (alliancecan.ca).
\end{acknowledgments}

\vspace{5mm}

\software{Gizmo~\citep{Hopkins_2015MNRAS.450...53H}
          }

\bibliography{sample631,inprep}{}
\bibliographystyle{aasjournal}

\end{document}